\documentclass[12pt]{article}
\usepackage{amsmath,amssymb,exscale,cancel}
\usepackage{slashed}
\usepackage{graphicx}
\usepackage{epsfig}
\usepackage{multicol}
\usepackage{color}
\usepackage{mathrsfs}
\usepackage{blindtext}
 \usepackage{fancyhdr}
\usepackage{hyperref}
\usepackage{cite}
\usepackage{mathtools}

\usepackage{floatrow}

\usepackage{graphicx}
\usepackage{sidecap}

\usepackage[latin1]{inputenc} 

\usepackage{multicol}  
 
 \usepackage{titlesec}
 
 \usepackage{rotating,slashed,xcolor,amsfonts,expdlist,charter}

\numberwithin{equation}{section}

\usepackage{xcolor}
\usepackage{sectsty}

\usepackage{mdframed}
\usepackage{titletoc}

\numberwithin{equation}{section}

\usepackage{xcolor}
\usepackage{sectsty}

\usepackage{mdframed}
\usepackage{titletoc}

\definecolor{secnum}{RGB}{13,151,225}
\definecolor{ptcbackground}{RGB}{212,237,252}
\definecolor{ptctitle}{RGB}{0,177,235}

\titlecontents{lsection}
  [5.8em]{\sffamily}
  {\color{secnum}\contentslabel{2.3em}\normalcolor}{}
  {\titlerule*[1000pc]{.}\contentspage\\\hspace*{-5.8em}\vspace*{5pt}%
    \color{white}\rule{\dimexpr\textwidth-15.5pt\relax}{1pt}}

\usepackage{hyperref}
\hypersetup{colorlinks,bookmarksopen,bookmarksnumbered,citecolor=rossos,
linkcolor=redy,pdfstartview=FitH,urlcolor=green-go}
\usepackage{slashed}

\definecolor{blus}{cmyk}{1,0.9,0,0.1}
\definecolor{verdes}{cmyk}{0.99,0,0.59,0.65}
\definecolor{rossos}{cmyk}{0,1,1,0.55}
\definecolor{redy}{cmyk}{0,1,1,0.7}
\definecolor{greeny}{cmyk}{0.99,0,0.59,0.98}
\definecolor{green-go}{cmyk}{0.79,0,0.59,0.5}

\usepackage{titlesec}

\def\tt{{\tiny \times}}

\newcommand{\beq}{\begin{equation}}
\newcommand{\eeq}{\end{equation}}

\def\hhref#1{\href{http://arxiv.org/abs/#1}{arXiv:#1}} 

 \def\Lag{\mathscr{L}}
 
\newcommand{\tmtextbf}[1]{{\bfseries{#1}}}
\newcommand{\tmtextrm}[1]{{\rmfamily{#1}}}
\newcommand{\bp}{\bar M_P}

\def\be{\begin{equation}}
\def\ee{\end{equation}}
\def\ba{\begin{array} }
\newcommand{\Tr}{\,{\rm Tr}}
\def\bac{\begin{array} {c}}
\def\bacc{\begin{array} {cc}}
\def\baccc{\begin{array} {ccc}}
\def\bacccc{\begin{array} {cccc}}
\def\ea{\end{array}}
\def\bea{\begin{eqnarray}}
\def\eea{\end{eqnarray}}

\definecolor{red}{rgb}{1,0,0}

\def\psl{\hbox{\hbox{${p}$}}\kern-1.9mm{\hbox{${/}$}}}
\def\dsl{\hbox{\hbox{${\partial}$}}\kern-2.2mm{\hbox{${/}$}}}
\def\Dsl{\hbox{\hbox{${D}$}}\kern-2.6mm{\hbox{${/}$}}}

\newcommand{\gappeq}{{\rlap{{\raise}.5ex\text{\ensuremath{>}}}{{\lower}.5ex\text{\ensuremath{\sim}}}}}
\newcommand{\lappeq}{{\rlap{{\raise}.5ex\text{\ensuremath{<}}}{{\lower}.5ex\text{\ensuremath{\sim}}}}}
\newcommand{\I}{\tmtextrm{1{\kern}-.24em l}}

\begin{document}
\topmargin -1.0cm
\oddsidemargin 0.9cm
\evensidemargin -0.5cm

{\vspace{-1cm}}
\begin{center}

\vspace{-1cm}

 {\LARGE \tmtextbf{ 
\color{rossos} \hspace{-0.9cm}    \\
Supercooling in Radiative Symmetry Breaking: Theory Extensions, Gravitational Wave Detection  \\ and Primordial Black Holes \hspace{-1.6cm}}} {\vspace{.5cm}}\\

\vspace{1.3cm}

{\large  {\bf Alberto Salvio }
{\em  
\vspace{.4cm}

 Physics Department, University of Rome Tor Vergata, \\ 
via della Ricerca Scientifica, I-00133 Rome, Italy\\

\vspace{0.6cm}

I. N. F. N. -  Rome Tor Vergata,\\
via della Ricerca Scientifica, I-00133 Rome, Italy\\

\vspace{.4cm}


\vspace{0.4cm}

\vspace{0.2cm}

 \vspace{0.5cm}
}

\vspace{.3cm}

}
\vspace{0.cm}

\end{center}

%
%
\noindent ---------------------------------------------------------------------------------------------------------------------------------
\begin{center}
{\bf \large Abstract}
\end{center}
\noindent  First-order phase transitions, which take place when the symmetries are predominantly broken (and masses are then generated) through radiative corrections, produce observable gravitational waves and primordial black holes. We provide a model-independent approach that is valid for large-enough supercooling to quantitatively describe these phenomena in terms of few parameters, which are computable once the model is specified. The validity of a previously-proposed approach of this sort is extended here to a larger class of theories. Among other things, we identify regions of the parameter space that correspond to the  background of gravitational waves recently detected by pulsar timing arrays (NANOGrav, CPTA, EPTA, PPTA) and others that are either excluded by the observing runs of LIGO and Virgo or within the reach of future gravitational wave detectors. Furthermore, we find regions  of the parameter space  where primordial black holes produced by large over-densities due to such phase transitions can account for  dark matter. Finally, it is shown how this model-independent approach can be applied to specific cases, including a phenomenological completion of the Standard Model with right-handed neutrinos and  gauged $B-L$ undergoing radiative symmetry breaking.

\vspace{0.7cm}

\noindent---------------------------------------------------------------------------------------------------------------------------------

  \vspace{-0.9cm}
  
  \newpage 
\tableofcontents

\noindent

\vspace{0.5cm}

\section{Introduction}\label{Introduction}

First-order phase transitions (PTs) leave potentially observable footprints that can be evidence for new physics  because the Standard Model (SM) does not feature this type of transitions. 

One example of such footprints is the spectrum of gravitational waves (GWs) produced by first-order PTs. GW astronomy has become an extremely active and exciting branch of physics after the discovery of the GWs from  binary black hole and neutron star mergers~\cite{Abbott:2016blz,TheLIGOScientific:2016wyq1,LIGOScientific:2017ync}. Very recently, the interest in this field has been further boosted by the detection of a background of GWs by pulsar timing arrays: these include the North American Nanohertz Observatory for Gravitational Waves (NANOGrav), the Chinese Pulsar Timing Array (CPTA), the European Pulsar Timing Array (EPTA) and the Parkes Pulsar Timing Array (PPTA)~\cite{NANOGrav:2023gor,Antoniadis:2023ott,Reardon:2023gzh,Xu:2023wog}. 

Another interesting consequence of first-order PTs is the production of primordial black holes (PBHs)~\cite{Hawking:1982ga,Crawford:1982yz,Kodama:1982sf,Moss:1994iq,Gross:2021qgx,Kawana:2021tde,Liu:2021svg,Baker:2021sno,Jung:2021mku,Hashino:2021qoq,Kawana:2022olo,Gouttenoire:2023naa}, which in turn can account for a fraction or the entire dark matter observed abundance.

First-order PTs always take place when the corresponding symmetry breaking is mostly induced (and masses are generated) radiatively, i.e.~through perturbative loop corrections~\cite{Salvio:2023qgb}. The seminal work on this radiative symmetry breaking (RSB) is Ref.~\cite{Coleman:1973jx} by Coleman and E.~Weinberg, which studied a simple toy model (see also Ref.~\cite{Levi:2022bzt}
for a recent analysis). The Coleman-Weinberg work was then extended to a more general field theory by Gildener and S.~Weinberg~\cite{Gildener:1976ih}. Furthermore, thanks to an approximate scale invariance, in the RSB scenario the first-order PTs feature a period of supercooling, when the temperature dropped by several orders of magnitude below the critical temperature~\cite{Witten:1980ez,Salvio:2023qgb}.

Supercooling in the RSB scenario allows us to be  confident about the validity of the one-loop approximation and the derivative expansion~\cite{Salvio:2023qgb}.
Moreover, it also ensures that the gravitational corrections to the false vacuum decay are amply negligible whenever the symmetry breaking scale  is small compared to the Planck mass, which is, of course, the most interesting case from the  phenomenological point of view.

Indeed, many RSB models featuring a strong first-order PT and predicting potentially observable GWs have been studied, ranging from electroweak (EW) symmetry breaking~\cite{Espinosa:2008kw,Farzinnia:2014yqa,Sannino:2015wka,Marzola:2017jzl,Brdar:2019qut,Kierkla:2022odc} to unified models~\cite{Huang:2020bbe}, passing through, e.g., Peccei-Quinn~\cite{Peccei:1977hh}   symmetry breaking~\cite{DelleRose:2019pgi,VonHarling:2019rgb,Salvio:2020prd,Ghoshal:2020vud} and the seesaw mechanism~\cite{Brdar:2018num,Kubo:2020fdd} (see Ref.~\cite{Salvio:2020axm} for a review). 

In~\cite{Salvio:2023qgb} it  was shown that a model-independent description of PTs and the consequent production of GWs in the RSB scenario is possible in terms of few parameters (which are computable once the model is specified) if enough supercooling occurred. Ref.~\cite{Salvio:2023qgb} provided a sufficient condition on the amount of  supercooling, which ensures that the model independent description  is valid. This led to a ``supercool expansion" in terms of a  quantity that is small when supercooling is large enough. 

In this work we investigate whether this condition can be weakened and, if so, how to systematically perform a corresponding ``extended supercool expansion". A weaker condition on supercooling is useful because it allows us to describe a larger class of models through the model-independent approach, without repeating the study of the PTs every time. 

Once one establishes that such extended supercool expansion can be performed, one can use it to describe in a model-independent way not only the spectrum of GWs, but also the production of PBHs due to the first-oder PTs. In particular, the amount of dark matter in the form of these PBHs can be determined in terms of the few parameters, which are computable once the model is specified. Moreover, one can identify the model-independent regions of parameter space corresponding to the GW background recently detected by pulsar timing arrays, as well as those excluded by the runs~\cite{KAGRA:2021kbb} of the Laser Interferometer Gravitational-Wave Observatory (LIGO)~\cite{Harry:2010zz,TheLIGOScientific:2014jea} and Advanced Virgo~\cite{VIRGO:2014yos}  and those within the reach of future GW detectors. These include the Laser Interferometer Space Antenna (LISA)~\cite{Audley:2017drz}, Cosmic Explorer (CE)~\cite{Evans:2016mbw,Reitze:2019iox}, 
Einstein Telescope (ET)~\cite{Punturo:2010zz, Hild:2010id, Sathyaprakash:2012jk}, the Big Bang Observer (BBO)~\cite{Crowder:2005nr, Corbin:2005ny, Harry:2006fi}, the Deci-hertz Interferometer Gravitational wave Observatory (DECIGO)~\cite{Seto:2001qf, Kawamura:2006}, etc.
 
  The paper is structured as follows.
  
 \begin{itemize}
 \item In Sec.~\ref{recap} we introduce the general class of theories featuring RSB, where the masses are mostly generated radiatively. These theories may include the SM or may be though of as ``dark" sectors weakly coupled to the SM. In the same section the model-independent description of RSB and the corresponding PTs in the supercool expansion is reviewed. This is necessary to render the subsequent original sections understandable and to establish our conventions.
 \item In Sec.~\ref{impro} we investigate when and how one can extend the validity of the model-independent description of PTs  to a larger class of RSB models by weakening the condition on the amount of supercooling.
 \item Sec.~\ref{Applications} is devoted to the possible applications of such extended supercool expansion to the production of GWs and PBHs through first-oder PTs. We also include in the discussion the  background of GWs recently discovered by pulsar timing arrays.
 \item Since the usefulness of the model-independent approach studied here is mainly due to the fact that one can avoid repeating the analysis of the PT in each RSB model, in Sec.~\ref{Examples of specific models} it is shown how to apply it to specific models by considering a couple of examples: a simple illustrative one and a phenomenological completion of the SM featuring right-handed neutrinos below the EW scale and the gauging of the difference $B-L$ between the baryon and lepton numbers, which undergoes RSB. In these examples the accuracy of the extended supercool expansion is also studied.
 \item Sec.~\ref{Conclusions} provides a detailed summary of the main original results of this paper and the final conclusions.
 \end{itemize}

\section{Supercool expansion: a recap}\label{recap}

In this section the important properties of RSB and the supercool expansion are summarized. This is necessary to explain in a clear way the original results of the subsequent sections. The reader can find in Ref.~\cite{Salvio:2023qgb} the proof of any non-trivial statement that is present in this section. We will also define our basic conventions here.

In the RSB scenario the sector responsible for the symmetry breaking is (at least approximately) classically scale invariant and it is thus described in general by a Lagrangian of the form
\be \label{eq:Lmatterns}
\Lag^{\rm ns}_{\rm matter} =  
- \frac14 F_{\mu\nu}^AF^{A\mu\nu} + \frac{D_\mu \phi_a \, D^\mu \phi_a}{2}  + \bar\psi_j i\slashed{D} \psi_j  - \frac12 (Y^a_{ij} \psi_i\psi_j \phi_a + \hbox{h.c.}) 
- V_{\rm ns}(\phi), 
\ee 
while gravity is assumed to be Einstein's gravity at the energies that are relevant for this work\footnote{It is possible, however, to construct a classically scale-invariant theory of gravity where scale invariance is broken by dimensional transmutation~\cite{Salvio:2014soa,Kannike:2015apa,Salvio:2017qkx,Salvio:2017xul,Salvio:2019wcp,Alvarez-Luna:2022hka} at energies that are assumed above those of interest here.}.
Here we consider generic numbers 
of real scalars $\phi_a$,   Weyl fermions $\psi_j$ and vectors $V^A_\mu$ (with field strength $F_{\mu\nu}^A$), respectively. The $V^A_\mu$ are gauge fields
and allow us to construct the covariant derivatives $D_\mu \phi_a$ and $D_\mu\psi_j$. Of course, in~(\ref{eq:Lmatterns}) all terms are gauge-invariant.
Also, the $Y^a_{ij}$  are the Yukawa couplings  and $V_{\rm ns}(\phi)$ has the general form
\be V_{\rm ns}(\phi)= \frac{\lambda_{abcd}}{4!} \phi_a\phi_b\phi_c\phi_d, \label{Vns}\ee
where $\lambda_{abcd}$ are the quartic couplings.

  In the RSB mechanism mass scales emerge radiatively from loops because there may be some specific energy 
at which the potential  in~(\ref{Vns}) develops a flat direction, $\phi_a = \nu_a \chi$, where $ \nu_a$ are the components of a unit vector $\nu$, i.e.~$ \nu_a  \nu_a =1$, and $\chi$ is a single scalar field. 
So, the RG-improved potential $V$ along $\nu$ reads
\be V(\chi) = \frac{\lambda_\chi (\mu)}{4}\chi^4, \qquad (\lambda_\chi(\mu) \equiv\frac1{3!} \lambda_{abcd}(\mu) \nu_a \nu_b \nu_c \nu_d). \label{Vvarphi}\ee 
Having a flat direction along $\nu$ for the RG energy $\mu$ equal to some specific value $\tilde\mu$ means $\lambda_\chi(\tilde\mu)=0$. Including the one-loop correction the quantum effective potential can be written
 \be V_q(\chi) = \frac{\bar \beta}4\left(\log\frac{\chi}{\chi_0}-\frac14\right)\chi^4,\label{CWpot}\ee
 where
 \be \bar\beta \equiv \left[\mu\frac{d\lambda_{\chi}}{d\mu}\right]_{\mu=\tilde\mu} \label{betabardef}\ee
 and $\chi_0$ is  related to $\tilde\mu$ through a renormalization-scheme-dependent formula. The field value $\chi_0$ is a stationary point of $V_q$ and, when $\bar\beta>0$, is also a point of minimum. Therefore, when the conditions 
  \beq\left\{
\begin{array}{rcll}
\lambda_\chi(\tilde\mu)  &=& 0 & \hbox{(flat direction),}\\
 & & \\ 
\left[\mu\frac{d\lambda_{\chi}}{d\mu}\right]_{\mu=\tilde\mu}  &>& 0 & \hbox{(minimum condition),}
\end{array}\right.
\label{eq:CWgen}
\eeq
 are satisfied one has a minimum of $V_q$ at a non-vanishing value $\chi_0$ of $\chi$ and the fluctuations of $\chi$ around $\chi_0$ have mass 
 \be m_\chi=\sqrt{\bar\beta}\, \chi_0. \label{mchi} \ee 
 
  This non-trivial minimum can generically break global and/or local symmetries and generate the  particle masses, with $\chi_0$ being the symmetry breaking scale. EW symmetry breaking can also be induced when there is a term in  $\Lag$ of the form
   \be  \Lag_{\chi h} = \frac12 \lambda_{\chi h}(\tilde\mu) \chi^2 |{\mathcal H}|^2,\label{portal} \ee  
 where ${\mathcal H}$ is the SM Higgs doublet and  $\lambda_{\chi h}$ is some coupling. Indeed,  by evaluating this term at $\chi=\chi_0$ we obtain the Higgs squared mass parameter
  \be \mu_h^2 = \frac12\lambda _{\chi h}(\tilde\mu) \chi_0^2. \label{muhInd}\ee
  So when $\lambda _{\chi h}(\tilde\mu) >0$ the masses of the SM elementary particles are generated. Recalling the well-known formula that relates $\mu_h^2$ and the Higgs mass, it is clear that we cannot use this mechanism to generate $\mu_h^2$ when $\chi_0$ is much below the EW scale and demand the validity of perturbation theory at the same time. Of course, it is still possible that the SM with an explicit scale-symmetry breaking parameter is weakly coupled to a scale-invariant sector that features RSB. In this case perturbation theory can be compatible with a $\chi_0$ much smaller than the EW scale.

Including now thermal corrections, the general expression of the effective potential $V_{\rm eff}$ at finite temperature $T$ is 
 (in the Landau gauge and at one-loop level\footnote{The assumption of supercooling allows us to trust the one-loop approximation~\cite{Salvio:2023qgb}. Resumming the daisy graphs~\cite{Arnold:1992rz} does not change this conclusion. Indeed, in order to keep perturbation theory valid in this resummation one should include not only the thermally induced masses~\cite{Kierkla:2023von} (a.k.a.~the thermal masses) but also the radiatively generated one, which is proportional to $\chi_0$. Therefore, all the daisy diagrams are suppressed by the large ratio $\chi_0/T$ due to supercooling.})
\be V_{\rm eff}(\chi,T) = V_q(\chi) +\frac{T^4}{2\pi^2}\left(\sum_b n_b J_B(m_b^2(\chi)/T^2)-2\sum_f J_F(m_f^2(\chi)/T^2)\right)+\Lambda_0,  \label{VeffSumm}  \ee
where the $m_b$ and $m_f$ are the background-dependent bosonic and fermionic masses, respectively, the sum over $b$ runs over all bosonic degrees of freedom and $n_b=1$ for a scalar (we work with real scalars) and $n_b=3$ for a vector degree of freedom.  In~(\ref{VeffSumm}) the sum over $f$, which runs over the fermion degrees of freedom, is multiplied by 2 because we work with Weyl spinors. Also,  
the thermal functions $J_B$ and $J_F$ are 
\bea \, \hspace{-1cm}J_B(x)\equiv \int_0^\infty dp\, p^2 \log\left(1-e^{-\sqrt{p^2+x}}\right)&=&-\frac{\pi^4}{45}+\frac{\pi^2}{12} x -\frac{\pi}{6} x^{3/2} -\frac{x^2}{32} \log\left(\frac{x}{a_B}\right) + O(x^3), \label{JBdef}\\
\hspace{-1cm}J_F(x)\equiv \int_0^\infty dp\, p^2 \log\left(1+e^{-\sqrt{p^2+x}}\right)&=& \frac{7\pi^4}{360}-\frac{\pi^2}{24} x -\frac{x^2}{32} \log\left(\frac{x}{a_F}\right) + O(x^3) \label{JFdef},\eea
 where $a_B = 16\pi^2 \exp(3/2-2\gamma_E)$, $a_F = \pi^2 \exp(3/2-2\gamma_E)$ and  $\gamma_E$ is the Euler-Mascheroni constant (the derivation of the expansions above are given in~\cite{Dolan:1973qd}). In Eq.~(\ref{VeffSumm}) we have included  a constant term $\Lambda_0$ to account for the observed value of the cosmological constant when $\chi$ is set to  the point of minimum of $V_{\rm eff}$. 
 
The PT associated with a radiative symmetry breaking  always turns out to be of first order. The absolute minimum of the effective potential is at $\chi=0$ for $T$ larger than the critical temperature $T_c$, while, for $T<T_c$, is at a non-zero temperature-dependent value. In the latter case the decay rate per unit of spacetime volume, $\Gamma$, of the false vacuum into the true vacuum can be computed with the formalism of~\cite{Coleman:1977py,Callan:1977pt,Linde:1980tt,Linde:1981zj}: 
\be  \Gamma\sim \exp(-S)\, ,\label{Gamma}\ee
where $S$ is the action
\be S=4\pi\int_0^{1/T} dt_E \int_0^\infty dr r^2\left(\frac12 \dot\chi^2+\frac12 \chi'^2+\bar V_{\rm eff}(\chi,T)\right),  \quad \bar V_{\rm eff}(\chi,T)\equiv V_{\rm eff}(\chi,T)-V_{\rm eff}(0,T)\label{SVeff}\ee
evaluated at the bounce, i.e.  the solution of the  differential problem~\cite{Salvio:2016mvj} 
\bea && \qquad \ddot\chi+\chi''+\frac{2}{r}\chi'= \frac{d\bar V_{\rm eff}}{d\chi}, \label{bounceProbE} \\ \dot\chi(r,0)=0, &&\,\dot\chi(r,\pm 1/(2T))=0, \quad  \chi'(0,t_E)=0, \quad \lim_{r\to \infty}\chi(r,t_E) = 0. \label{bounceProb}\eea
A dot and a prime denote a derivative with respect to the Euclidean time $t_E$ and the spatial radius $r\equiv \sqrt{\vec{x}^{\,2}}$, respectively.  A particular solution of~(\ref{bounceProbE})-(\ref{bounceProb}) is the time-independent bounce, 
\be \chi''+\frac{2}{r}\chi'= \frac{d\bar V_{\rm eff}}{d\chi}, \qquad \chi'(0)=0, \quad \lim_{r\to \infty}\chi(r) = 0, \label{bounceProb3}\ee
for which
\be S =\frac{S_3}{T}, \qquad  \quad S_3 \equiv 4\pi \int_0^\infty dr \, r^2\left(\frac12 \chi'^2+\bar V_{\rm eff}(\chi,T)\right). \label{S3f}\ee
If the time-independent bounce dominates, the decay rate is~\cite{Linde:1980tt,Linde:1981zj}
\be  \Gamma\approx T^4\left(\frac{S_3}{2\pi T}\right)^{3/2}\exp(-S_3/T) \label{Gamma3}\ee
and $S_3$ evaluated at the time-independent bounce can be written as follows:
\be S_3 =  -8\pi \int_0^\infty dr \, r^{2} \bar V_{\rm eff}(\chi,T). \label{bounceSd}\ee

 As long as perturbation theory holds, in a generic theory with RSB, Eq.~(\ref{eq:Lmatterns}),  when $T$ goes below $T_c$ the scalar field $\chi$ is always trapped in the false vacuum  until  $T$ is much below $T_c$, i.e. the universe always features a phase of supercooling.  If this process  is strong enough, in a generic theory of the form~(\ref{eq:Lmatterns}) the full effective action for relevant values of $\chi$ can be described by three and only three parameters: $\chi_0$, $\bar\beta$ and a real, non-negative and $\chi$-independent  quantity $g$,
 \be g^2 \equiv \sum_b n_bm_b^2(\chi)/\chi^2+\sum_f m^2_f(\chi)/\chi^2, 
\label{M2g2def}\ee
which plays the role of a ``collective coupling" of $\chi$ with all fields of the theory. This is possible because the field value $\chi_b$ around the barrier, which can be defined by $\bar V_{\rm eff}(\chi_b,T)  = 0$, turns out to be small compared to $T$ for  large-enough supercooling:
\be \frac{\chi_b^2}{T^2}  \approx \frac{g^2}{6\bar\beta \log\frac{\chi_0}{T}}, \label{chibTorder} \ee
such that the small-field expansions in~(\ref{JBdef}) and~(\ref{JFdef}) can be truncated as 
\bea J_B(x) &\approx& J_B(0)+\frac{\pi^2}{12} x ,\label{JBapp}  \\
J_F(x) &\approx&J_F(0)-\frac{\pi^2}{24}x, \label{JFapp}\eea
 and the logarithmic term  in $V_q$  can be written as follows:
\be \log\frac{\chi_b}{\chi_0} -\frac14= \log\frac{\chi_b}{T}-\frac14+\log\frac{T}{\chi_0}\approx \log\frac{T}{\chi_0}. \label{logApp}\ee 
A sufficient condition for the approximations in~(\ref{JBapp}) and~(\ref{JFapp}) to be valid is that $\epsilon$ is small, where 
\be \epsilon\equiv  \frac{g^4}{6\bar\beta \log\frac{\chi_0}{T}}
 \label{CondConv}\ee
 Using now the approximations in~(\ref{logApp}),~(\ref{JBapp}) and~(\ref{JFapp}), the bounce action can be computed with 
\be \bar V_{\rm eff}(\chi,T) \approx \frac{m^2(T)}{2} \chi^2-\frac{\lambda(T)}{4} \chi^4 \label{barVapp}\ee
where $m$ and $\lambda$ are real and positive functions of $T$ defined by
\be m^2(T) \equiv \frac{g^2 T^2}{12}, \qquad \lambda(T) \equiv \bar\beta \log\frac{\chi_0}{T}. \label{mlambdaDef}\ee
For this effective potential the tunneling process is dominated by the time-independent bounce.
The bounce action $S_3$ computed with the effective potential in~(\ref{barVapp}) turns out to be
\be S_3=c_3 \frac{m}{\lambda}, \qquad c_3=18.8973...\label{S3c3}\ee 
 (see also~\cite{Brezin-Parisi,Arnold:1991cv} for previous calculations).

In general the nucleation temperature $T_n$ can be defined as the temperature for which $\Gamma/H_I^4= 1$, so, using the fact that the decay is dominated by the time-independent bounce, at $T=T_n$
\be  \frac{S_3}{T_n}-\frac32 \log \left(\frac{S_3/T_n}{2\pi}\right) \approx 4 \log \left(\frac{T_n}{H_I}\right),\label{TnEq0}\ee
where 
\be H_I = \frac{\sqrt{\bar\beta} \chi_0^2}{4\sqrt{3}\bp}\ee
is the Hubble rate associated with the exponential expansion of space that takes place during supercooling. By using the expression of $S_3$ in~(\ref{S3c3}) and the definitions in~(\ref{mlambdaDef}) one finds the following solution for $T_n$
\be X\equiv \log\frac{\chi_0}{T_n} \approx \frac{c-\sqrt{c^2-16a}}8, \label{appTn}\ee
with
\be a\equiv \frac{c_3g}{\sqrt{12}\bar\beta}, \quad c\equiv 4\log\frac{4\sqrt{3}\bp}{\sqrt{\bar\beta}\,\chi_0}+\frac32\log\frac{a}{2\pi} \label{caDef}\ee
and $\bp$ is the reduced Planck mass.

In general, the strength of the PT is measured by the parameter $\alpha$ defined as~\cite{Caprini:2019egz,Ellis:2019oqb}
\be \alpha \equiv \frac{30 \rho(T_n)}{\pi^2 g_*(T_n)T_n^4}, \ee
where $g_*(T)$ is the effective number of relativistic species at temperature $T$, in the presence of supercooling
 \be \rho(T_n) \approx \left[-\bar V_{\rm eff}(\langle\chi\rangle,T)\right]_{T=T_n}  \ee
 and $\langle\chi\rangle$ is the point of absolute minimum of the full effective potential. For an RSB PT $\alpha \gg 1$.

 Another important parameter to analyse the production of  GWs and PBHs  is the inverse duration $\beta$ of the PT that, in models with supercooling, is~\cite{Caprini:2015zlo,Caprini:2018mtu,VonHarling:2019rgb} 
  \be\beta = \left[\frac1{\Gamma}\frac{d\Gamma}{dt}\right]_{t_n},\ee 
  where   $t_n$ is the value of the time $t$ when $T=T_n$. Recalling 
that the tunneling process is dominated by the time-independent bounce,
\be \beta \approx H_n\left[T\frac{d}{dT}(S_3/T)-4-\frac{3}{2}T\frac{d}{dT}\log(S_3/T)\right]_{T=T_n}, \label{betaH1}\ee 
where $H_n\approx H_I$ is the Hubble rate  when $T=T_n$. 
 
 Note that here we are relying on a small $\epsilon$ expansion (a ``supercool expansion") and what we have done so far is the analysis at leading order (LO), that is modulo terms of relative order $\sqrt{\epsilon}$. Including these terms and  treating them perturbatively would mean working in the supercool expansion at next-to-leading order (NLO). This can be done by including the term of order $x^{3/2}$ in the expansion of $J_B(x)$, Eq.~(\ref{JBdef}), and is justified if $\epsilon$ is small. In Sec.~\ref{impro} we will explain how to extend the supercool expansion to order-one values of $\epsilon$.

The effective potential at NLO, therefore, includes a  cubic-in-$\chi$ term and reads
\be \bar V_{\rm eff}(\chi,T) \approx \frac{m^2(T)}{2} \chi^2-\frac{k(T)}{3}\chi^3-\frac{\lambda(T)}{4} \chi^4, \label{barVnlo}\ee
where $m^2$ and $\lambda$ are defined 
in~(\ref{mlambdaDef}), 
\be k(T)\equiv \frac{\tilde g^3 T}{4\pi}, \label{kdef}\ee
and $\tilde g$ is a  real, non-negative and $\chi$-independent parameter  defined by
\be \tilde g^3 \equiv \sum_b n_bm_b^3(\chi)/\chi^3. \label{gtdef}\ee
 This is an extra parameter that is needed for a model-independent description of this scenario at NLO. In general we have
\be \tilde g\leq g. \label{disggt}\ee

To understand why the term cubic in $\chi$ in~(\ref{barVnlo}) can be considered as a small correction in the supercool expansion, one can rescale $\chi\to \chi/\sqrt{\lambda}$ in the bounce action, Eq.~(\ref{SVeff}), to obtain 
\be S = \frac{4\pi}{\lambda}\int_0^{1/T} dt_E \left[ \int_0^\infty dr \, r^2\left(\frac12 \dot\chi^2+\frac12 \chi'^2+\frac{m^2}{2} \chi^2-\frac{1}{4} \chi^4\right) -\frac{k}{3\sqrt{\lambda}} \int_0^\infty dr \, r^2\chi^3 \right].\label{SNLO}\ee 
Since we  eventually need to set $T=T_n$,  the term proportional to $k$ has relative order at most $\sqrt{\epsilon}$ times a small number $\approx 1/(\sqrt{2}\pi)$ (where the LO result $S_{3}\approx 4\pi gT/(\sqrt{12}\lambda)$ has been used).
Working with the supercool expansion at NLO (i.e. treating the cubic term in~(\ref{barVnlo}) perturbatively at first order) one can then find corrected analytical expressions for $T_n$, $S_3$ and $\beta$, which depend on the extra parameter $\tilde g$. For example, 
\be S_3 = \frac{1}{\lambda}\left(c_3 m -\tilde c_3\frac{k}{3\sqrt{\lambda}}\right), \label{S3k0}\ee
where 
\be \tilde c_3 \equiv 4\pi \int_0^\infty  dr \,  r^2\chi_{\rm LO}^3 \ee
and $\chi_{\rm LO}$ is the LO bounce configuration.  Of course, one can then go ahead and compute smaller and smaller corrections.
 
 \section{Extending the validity of the  supercool expansion}\label{impro}
  
In this section we study when and how one can extend the validity of the supercool expansion to cases in which 
\be \epsilon \sim 1. \ee

\subsection{Several degrees of freedom}\label{Several degrees of freedom}

The expansion developed in~\cite{Salvio:2023qgb}, which we have reviewed in Sec.~\ref{recap}, generally works for $\epsilon$ small. However, it also holds for values of $\epsilon$ of order one if there are several degrees of freedom, say $N$, with dominant couplings (all of the same order of magnitude, say $\tau$) to the flat-direction field $\chi$. 
Indeed,   in this case $g$ defined in~(\ref{M2g2def}) scales as $g \sim \sqrt{N}\tau$, while $\tilde g$ defined in~(\ref{gtdef}) scales as $\tilde g \lesssim \sqrt[3]{N}\tau$, and so 
$\tilde g^3/g^3 \lesssim1/\sqrt{N}$: the inequality here is due to the fact that $\tilde g$ receives contributions only from bosons, while both fermions and bosons contribute to $g$. As a result, the extra cubic term in the bounce action of Eq.~(\ref{SNLO}) gets a further suppression factor (see Eq.~(\ref{kdef})), which is at least as small as $1/\sqrt{N}$. On the other hand, 
\begin{itemize}
\item since $1/X = 6\bar\beta\epsilon/g^4$, for order one $\epsilon$ the quantity $1/X$ is still small because $\bar\beta$ is loop suppressed and so the approximation in~(\ref{logApp}) is still good,
\item truncating the small-$x$ expansions in~(\ref{JBdef}) and~(\ref{JFdef}) up to the $x^{3/2}$ term is still justified because the higher-order terms involve smaller and smaller coefficients\footnote{One can check that by looking at the full expansions of $J_B(x)$ and $J_F(x)$ provided, for example, in~\cite{Quiros:1994dr}.}, with the coefficient of the $O(x^2)$ term being already quite small, $\sim 1/32$. 
\end{itemize}

\subsection{Improved supercool expansion}\label{Improved supercool expansion}

On the other hand, if the number of degrees of freedom with a dominant coupling to $\chi$ is too small, one instead has $\tilde g\approx g$ and, in this case, the expansion of Sec.~\ref{recap} breaks down for order 1 values of $\epsilon$ (although it still holds for small $\epsilon$).

\subsubsection{Bounce solution and action}

 In order to extend the class of theories that can be described by the supercool expansion one is, therefore, interested in including the cubic term in~(\ref{barVnlo}) in the non-perturbative computation of the bounce action and treating the other corrections as perturbations (indeed, they are still small as long as $\epsilon$ is at most of order one, as we have  seen in Sec.~\ref{Several degrees of freedom}). We will refer to this improvement as the ``improved supercool expansion". Let us explain how to construct it.

The expression of $\bar V_{\rm eff}$ in~(\ref{barVnlo}), together with the form of the bounce problem in~(\ref{bounceProbE})-(\ref{bounceProb}), tells us that the characteristic bounce size $R_b$ is of order $R_b\sim 1/m(T)\gtrsim1/T$, where in the second estimate we have used the perturbativity condition that $g$ is not too large. Indeed, the bounce size can be read from the large-$r$ limit of the bounce solution and in this limit the last condition in~(\ref{bounceProb}) tells us that only the quadratic term in~(\ref{barVnlo}) matters. Therefore, the bounce solutions  are approximately time-independent even including the cubic term in~(\ref{barVnlo}).

\begin{figure}[t]
\begin{center}
  \includegraphics[scale=0.48]{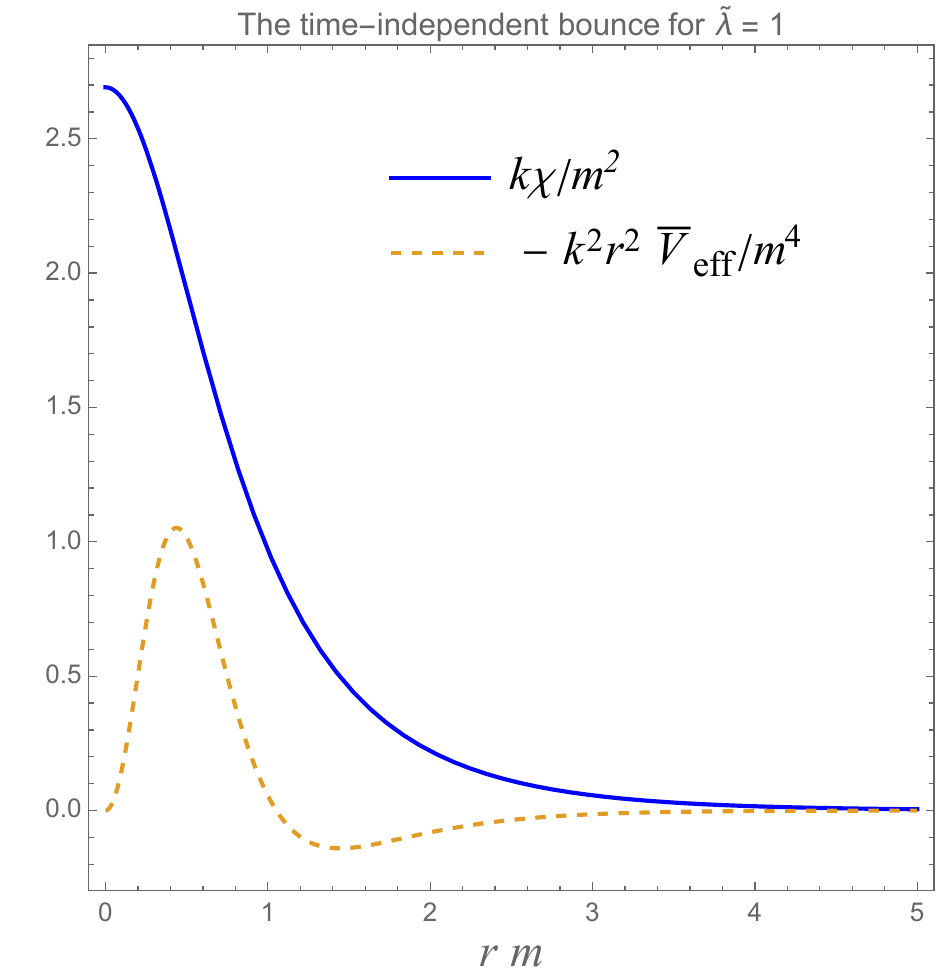}  \hspace{1cm} \includegraphics[scale=0.48]{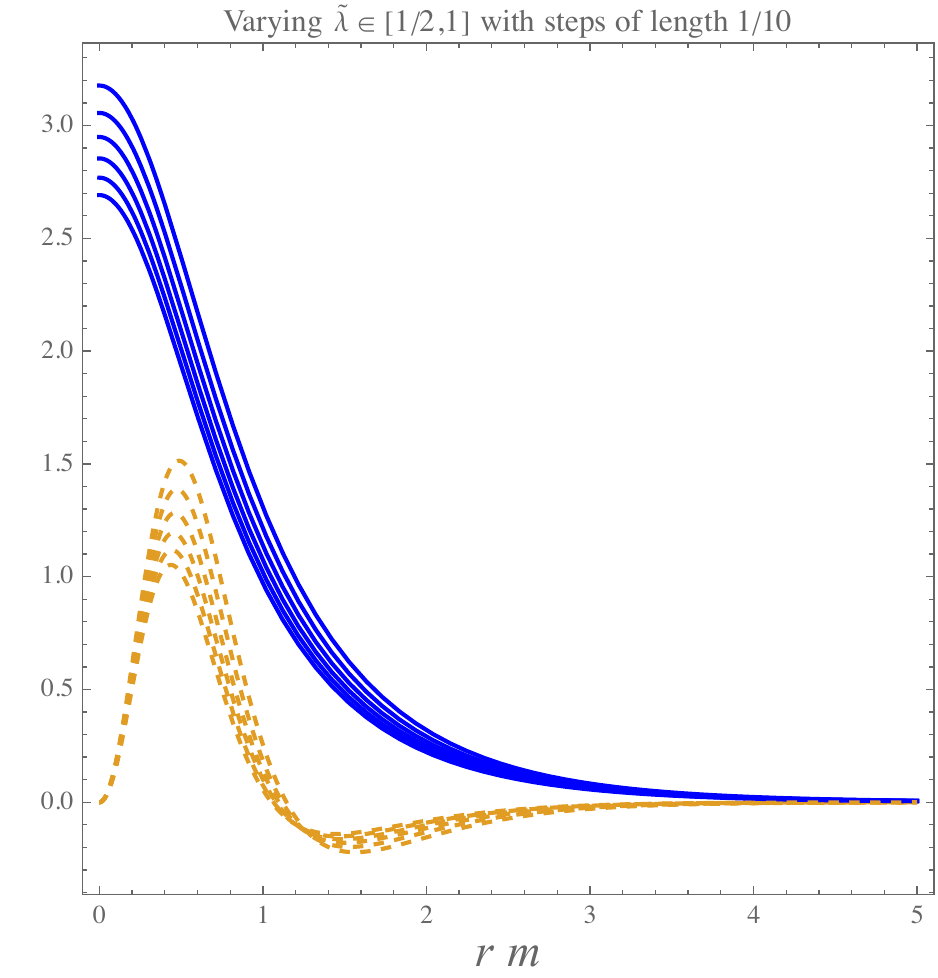}  
    \caption{\em  The relevant bounce and the corresponding integrand function (divided by $8\pi l\xi^2$) appearing in the bounce action, Eq.~(\ref{S3improS}), for the effective potential (\ref{barVnlo}) and varying $\tilde \lambda\equiv  \lambda m^2/k^2$.}\label{kappasetting}
  \end{center}
\end{figure}

Looking then at~(\ref{S3f}) and redefining~\cite{Levi:2022bzt} $r\equiv l\rho$  and $\chi\equiv \xi \varphi$ one obtains the bounce action for the new radial variable $\rho$ and the new field $\varphi$ 
\be S_3 \equiv 4\pi l\xi^2\int_0^\infty d\rho \, \rho^2\left(\frac12 \left(\frac{d\varphi}{d\rho}\right)^2+\tilde V_{\rm eff}(\varphi,T)\right), \ee 
where
\be \tilde V_{\rm eff}(\varphi,T)\equiv \left(\frac{l}{\xi}\right)^2 \bar V_{\rm eff}(\chi,T).\ee
By evaluating at the bounce solution one then obtains, like in~(\ref{bounceSd}), a simplified bounce action
\be S_3 =  -8\pi l \xi^2 \int_0^\infty d\rho \, \rho^{2} \tilde V_{\rm eff}(\varphi,T). \label{S3improS}\ee
Choosing now
\be l = \frac1{m}, \qquad \xi = \frac{m^2}{k}, \ee
with $m$ and $k$ defined in~(\ref{mlambdaDef}) and~(\ref{kdef}), respectively, gives
\bea \tilde V_{\rm eff}(\varphi,T) &=& \frac12 \varphi^2-\frac13 \varphi^3 -\frac{\tilde \lambda}{4}\varphi^4, \\
S_3 &=&-\frac{8\pi m^3}{k^2}\int_0^\infty d\rho \, \rho^{2} \left( \frac12 \varphi^2-\frac13 \varphi^3 -\frac{\tilde \lambda}{4}\varphi^4\right)  \eea
where
\be \tilde \lambda \equiv  \frac{\lambda m^2}{k^2} >0 \label{deflambdat} \ee 
and $\lambda$ defined in~(\ref{mlambdaDef}).
The quantity $\tilde \lambda$ can also be rewritten by using~(\ref{mlambdaDef}) and~(\ref{kdef}) as follows
\be \tilde \lambda(T) =  \frac{(4\pi)^2 \bar\beta }{12\, \tilde g^6/g^2} \log(\chi_0/T), \ee 
which depends on $T$ only through $\log(\chi_0/T)$.  Using the definition of $\epsilon$ in~(\ref{CondConv}) one obtains  
\be \tilde \lambda = \frac{2\pi^2}{9\epsilon} \frac{g^6}{\tilde g^6}, \label{lambda-eps}\ee
and recalling the bound in~(\ref{disggt})
\be \tilde \lambda \geq \frac{2\pi^2}{9\epsilon}.\ee
So the small-$\epsilon$ expansion of Sec.~\ref{recap} corresponds to $\tilde \lambda$ large. Here we are interested in setting $\epsilon$ of order 1 and $\tilde g \approx g$, when that expansion breaks down. Thus we are interested in finite values of $\tilde \lambda$ around $1$. 
In Fig.~\ref{kappasetting} the time-independent bounces for $\tilde \lambda\in[1/2,1]$ are shown, together with  $-\rho^{2} \tilde V_{\rm eff}$, which appears in the integrand of the bounce action in~(\ref{S3improS}).

We are not able to find the analytic dependence of the bounce action $S_3$ on $\kappa$. However, one can compute the bounce and the corresponding $S_3$ for several values of $\kappa$ and then perform a fit~\cite{Adams:1993zs,Sarid:1998sn,Levi:2022bzt}. Doing so we find that
\be S_3 =  \frac{27\pi m^3}{2k^2} \frac{1+\exp(-1/\sqrt{\tilde \lambda})}{1+\frac92 \tilde \lambda} =27\pi m^3 \frac{1+\exp(-k/(m\sqrt{\lambda}))}{2k^2+9\lambda m^2} \label{S3k} \ee 
reproduces the numerical calculations at the $\sim 1\%$ level for the values of $\tilde \lambda$ we are interested in. The result in~(\ref{S3k}) was found by~\cite{Levi:2022bzt} in a specific setup.
Here its validity has been established in a model-independent way within the improved supercool expansion.

\subsubsection{Nucleation temperature}\label{Tnimpro}

\begin{figure}[t]
\begin{center}
\hspace{-0.4cm}  
\includegraphics[scale=0.67]{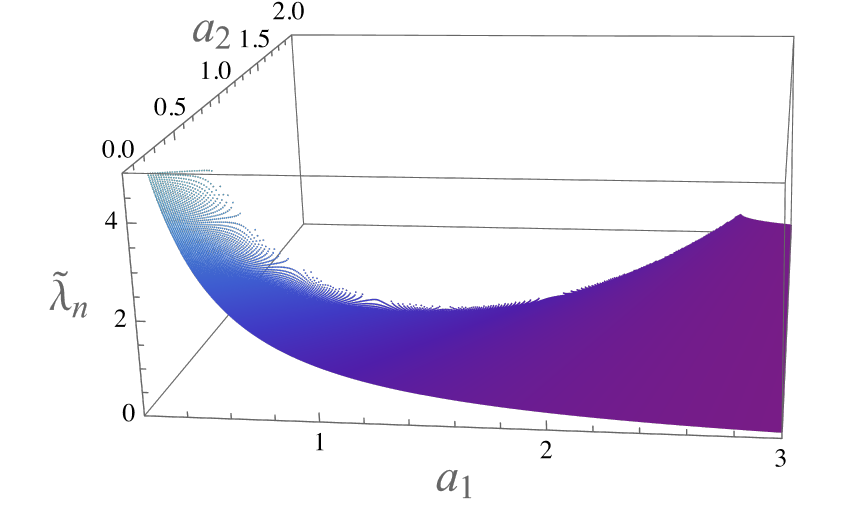}  \hspace{0.5cm}
 \includegraphics[scale=0.43]{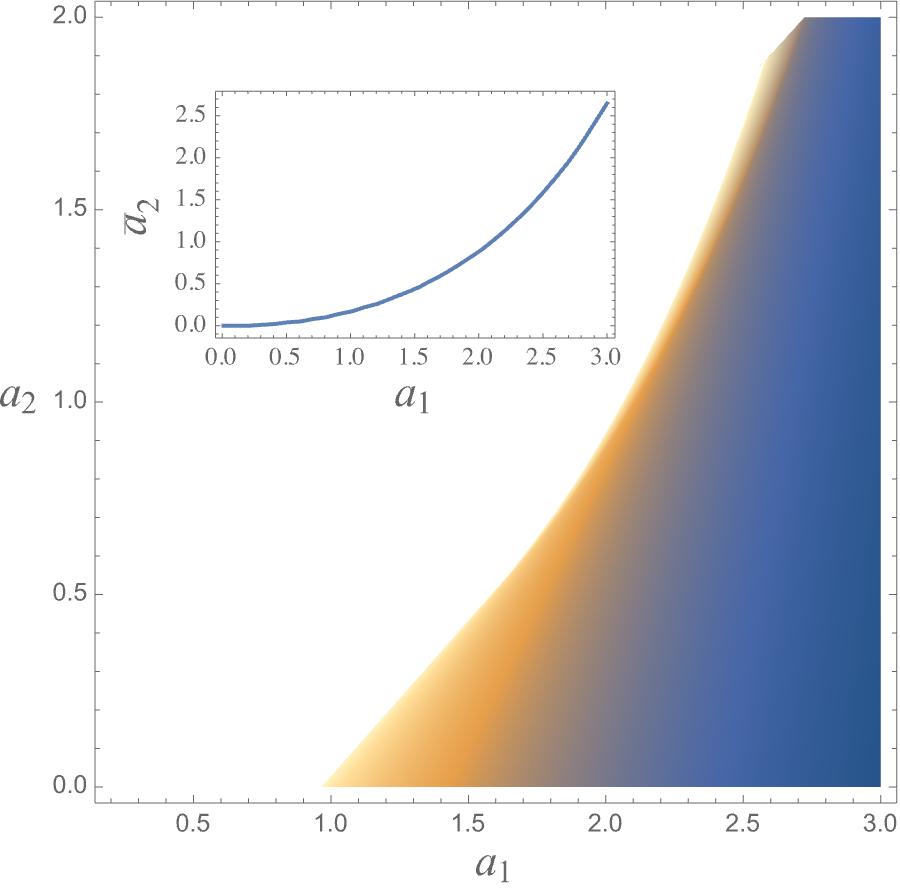}\includegraphics[scale=0.50]{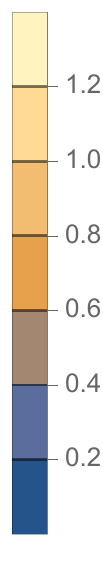}  
    \caption{\em The  solution $\tilde \lambda_n$ of Eq.~(\ref{lteq}) as a function of $a_1$ and $a_2$ defined in~(\ref{a1a2}). The inset in the right plot gives the maximal value of $a_2$ for a given $a_1$ such that the  solution $\tilde \lambda_n$ exists. Using the definitions of $\tilde \lambda$ and $\lambda$ in (\ref{deflambdat}) and (\ref{mlambdaDef}) one can extract the nucleation temperature.}
    \label{lambdat}
  \end{center}
\end{figure}

Inserting the expression in~(\ref{S3k}) into the equation for the nucleation temperature $T_n$ in~(\ref{TnEq0}) leads to a complicated non-polynomial equation in $\tilde \lambda$. This equation can be partially simplified by dropping the second term on the left-hand side of Eq.~(\ref{TnEq0}), which is always negligible with respect to the first one because the semiclassical approximation requires $S_3/T$ large. Within this approximation the equation for $\tilde \lambda$ reads 
\be a_1-a_2 \tilde \lambda = F(\tilde \lambda) \equiv \frac{1+\exp(-1/\sqrt{\tilde \lambda})}{2/9+\tilde\lambda}, \label{lteq} \ee
where 
\be a_1 \equiv \frac{c\,  c_3 k^2}{3\pi a\, \bar\beta\, m^2}, \qquad a_2 \equiv \frac{4 c_3 k^4}{3\pi a\,  \bar\beta^2m^4}, \label{a1a2}\ee 
the value of $c_3$ is given in~(\ref{S3c3}) and $a$ and $c$ are defined in Eq.~(\ref{caDef}) (the  term $\frac32\log\frac{a}{2\pi}$ in $c$ can be dropped as it comes from the second term on the left-hand side of Eq.~(\ref{TnEq0})). Here we are interested in the smallest real and positive solution $\tilde \lambda_n\equiv \tilde\lambda(T_n)$ of Eq.~(\ref{lteq}) for which the straight line $a_1-a_2\tilde \lambda$ reaches $F(\tilde \lambda)$ from below in increasing $\tilde \lambda$ (that corresponds to $\Gamma$ reaching $H_I^4$ from below). 
Clearly, such a solution does not always exist for any $a_1$ and $a_2$. First, one must have $a_1\leq F(0) =9/2$; second, for each given $a_1$ the parameter $a_2$ must me smaller than a certain critical value $\bar a_2(a_1)$, which is given in the inset of the right plot of Fig.~\ref{lambdat}.  Fig.~\ref{lambdat} also shows as a function of $a_1$ and $a_2$ the solution $\tilde\lambda_n$ (when it exists), which has been  obtained numerically. Tables containing the numerical determination of $\bar a_2$ as a function of $a_1$ and of $\tilde \lambda_n$ as a function of $a_1$ and $a_2$ can be found at \cite{dataset}.  Once we fix the parameters $g$, $\bar\beta$, $\chi_0$ and $\tilde g$ the quantities $a_1$ and $a_2$ as well as $\tilde \lambda_n$ and thus the nucleation temperature $T_n$ are fixed.  
%
%

Using the obtained solution $\tilde \lambda_n$ we checked that the PT strength parameter $\alpha$ is large in an RSB PT  for realistic and perturbative values of the parameters, even in the improved supercool expansion discussed in this paper. Thus, the plasma effects (such as those studied in Refs.~\cite{Bodeker:2009qy,Bodeker:2017cim}) can be neglected in this particular scenario.

 One might wonder whether the effect of the spacetime curvature due to $H_I\neq 0$ can alter the decay rate. In standard Einstein gravity, this may happen if $T_n$ is so small to be comparable with $H_I$. We checked that, whenever a solution for $\tilde\lambda_n$ exists, this never happens, at least for realistic and perturbative values of the parameters. On the other hand, if a solution for $\tilde\lambda_n$ does not exist,  the effect of the spacetime curvature, along with quantum fluctuations, can eventually become important in the decay rate~\cite{Kearney:2015vba,Joti:2017fwe,Markkanen:2018pdo,DelleRose:2019pgi} and lead to the completion of the transition.
 
 \subsubsection{Duration of the phase transition}\label{Duration of the phase transition}
 
 Using the expression of $S_3$ in~(\ref{S3k})  and dropping the last term in~(\ref{betaH1}), which is negligible in the semiclassical approximation as we have pointed out in Sec.~\ref{Tnimpro}, we obtain a formula for the inverse duration of the PT:
 \be \frac{\beta}{H_n} \approx \frac{\pi^3g^5}{6\sqrt{3}\tilde g^8}  \frac{(4\pi)^2\bar\beta}{\tilde g^4}(-F'(\tilde \lambda_n)) -4, \ee
 where $F'$ is the derivative of $F$ defined in Eq.~(\ref{lteq}) with respect to $\tilde \lambda$; note that $F$ is a monotonic decreasing function of $\tilde \lambda$ so $-F'>0$. 
 
\begin{figure}[t!]
\begin{center}
\hspace{-0.4cm}  
\includegraphics[scale=0.47]{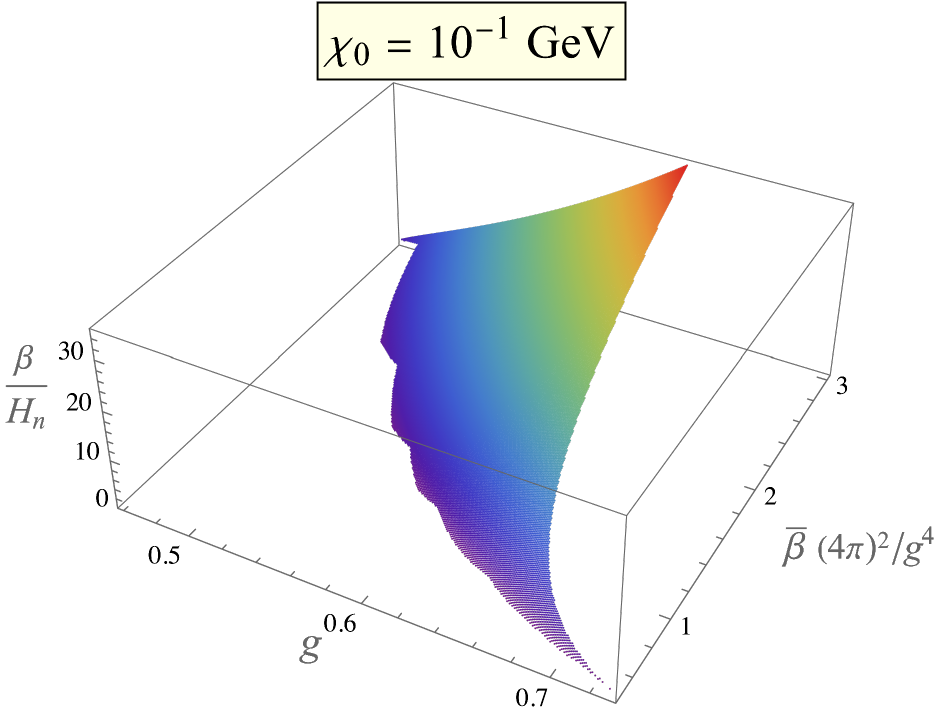}  \hspace{0.3cm}
 \includegraphics[scale=0.38]{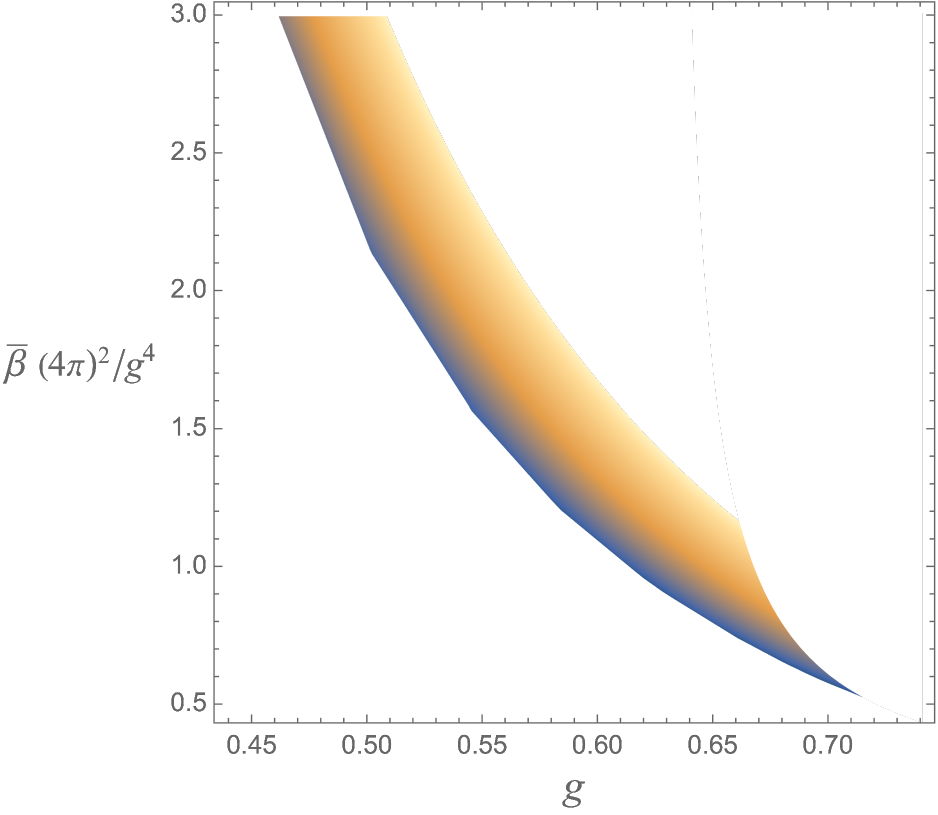}\includegraphics[scale=0.40]{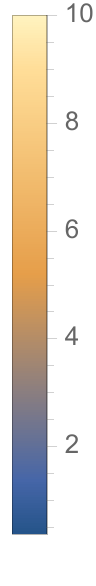}

\includegraphics[scale=0.47]{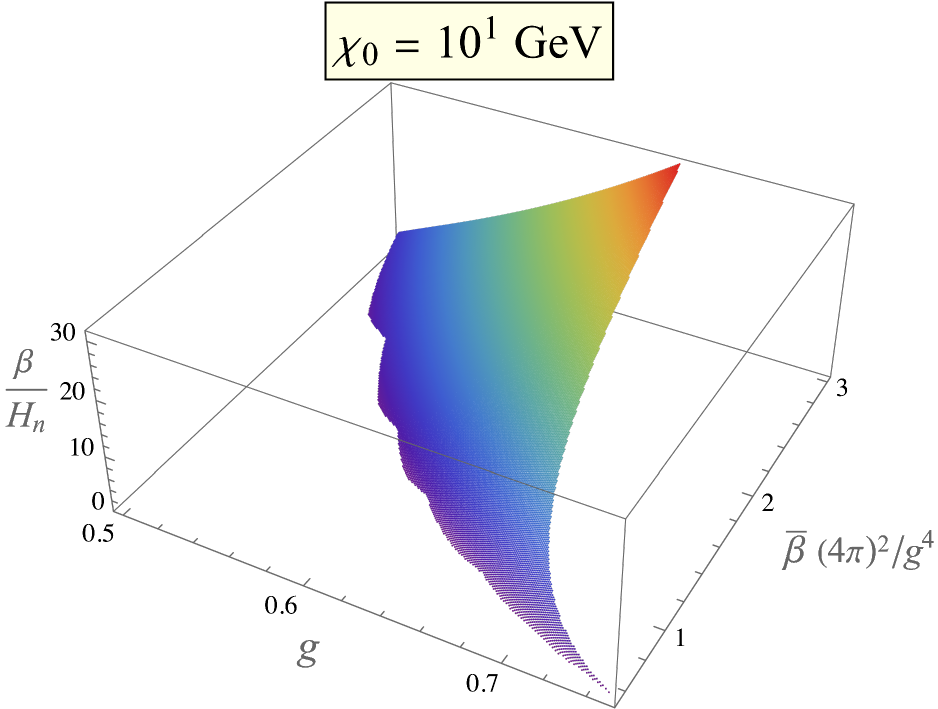}  \hspace{0.3cm}
 \includegraphics[scale=0.38]{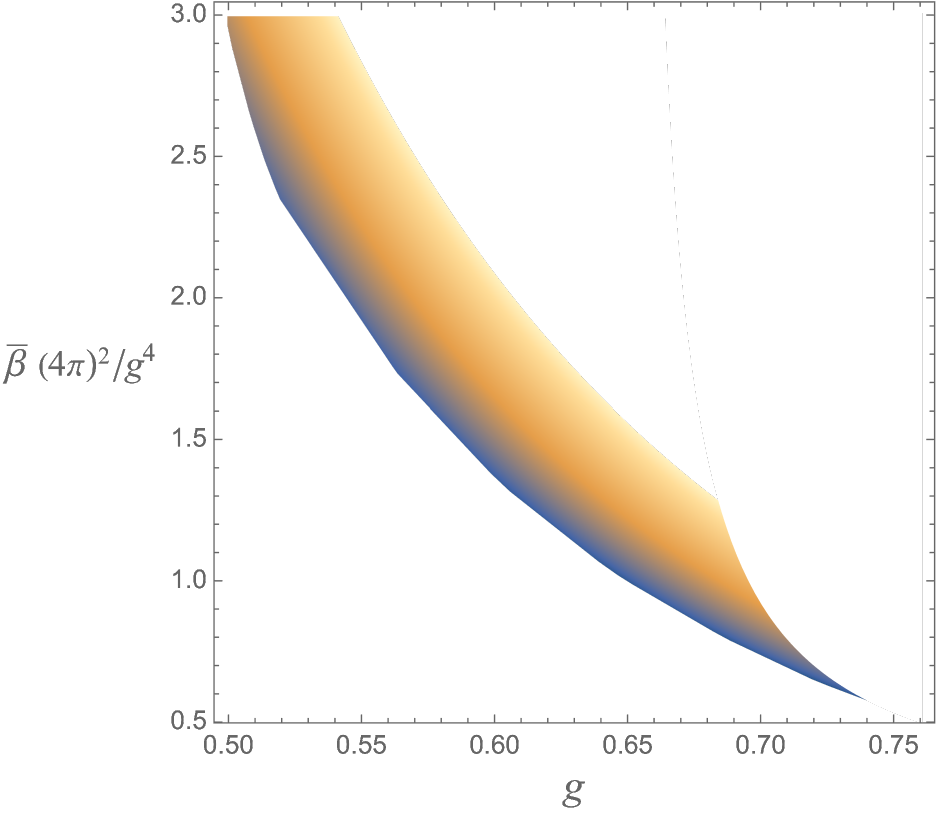}\includegraphics[scale=0.40]{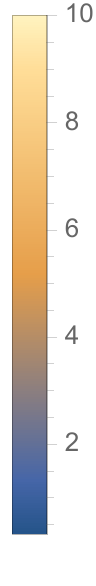}
 
\includegraphics[scale=0.47]{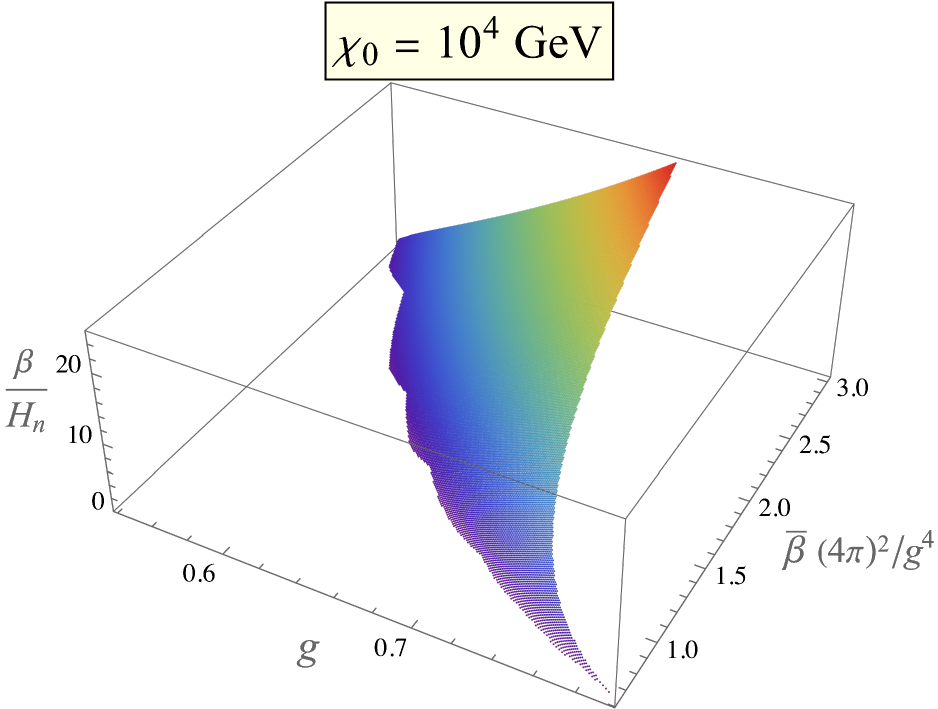}  \hspace{0.3cm}
 \includegraphics[scale=0.38]{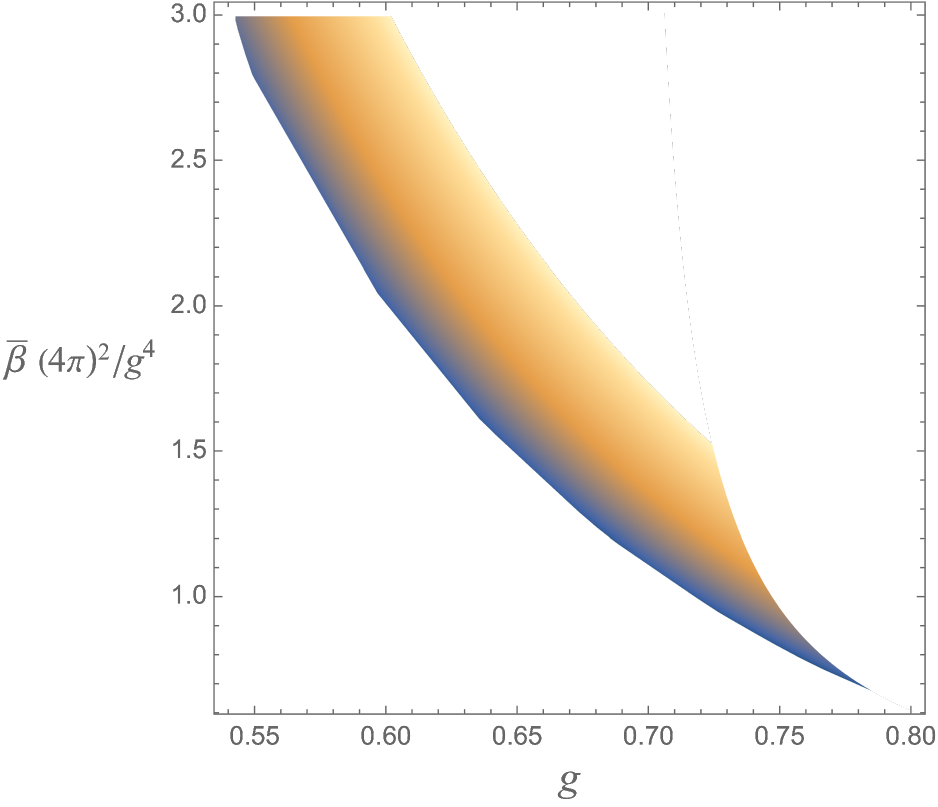}\includegraphics[scale=0.40]{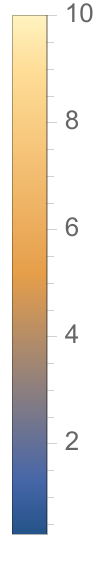}
   \vspace{-0.4cm}
    \caption{\em Inverse duration $\beta$ of the phase transition in units of the Hubble rate $H_n$ as a function of $g$ and $\bar\beta$ for various values of the symmetry breaking scale $\chi_0$. Here $\tilde g = g$ and $\epsilon<3$ has been imposed to guarantee the validity of  the improved supercool expansion.}
    \label{betagb}
  \end{center}
\end{figure}

\begin{figure}[t!]
\begin{center}
\includegraphics[scale=0.47]{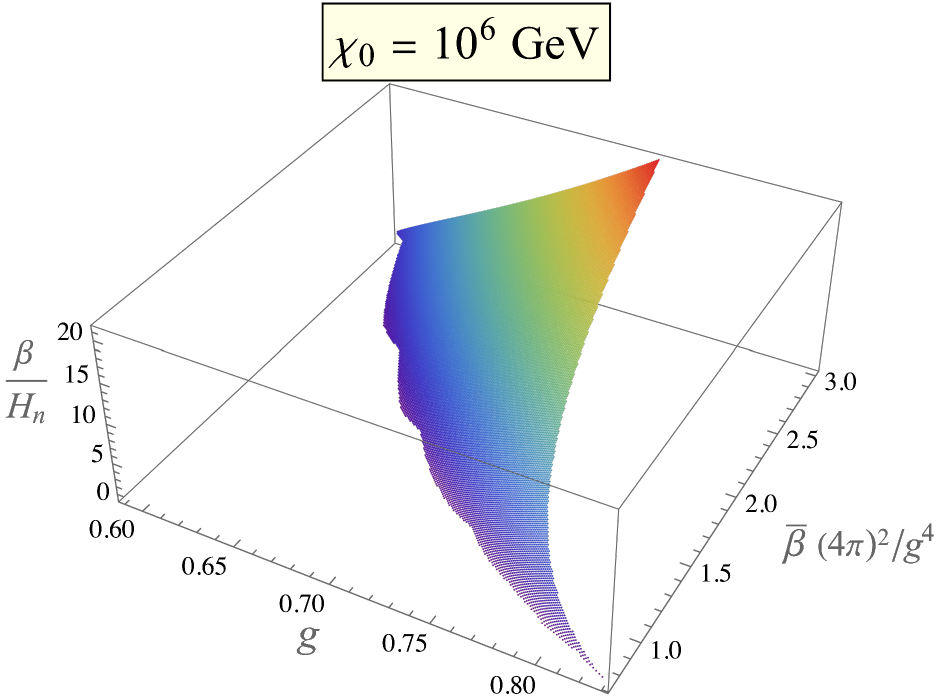}  \hspace{0.3cm}
 \includegraphics[scale=0.38]{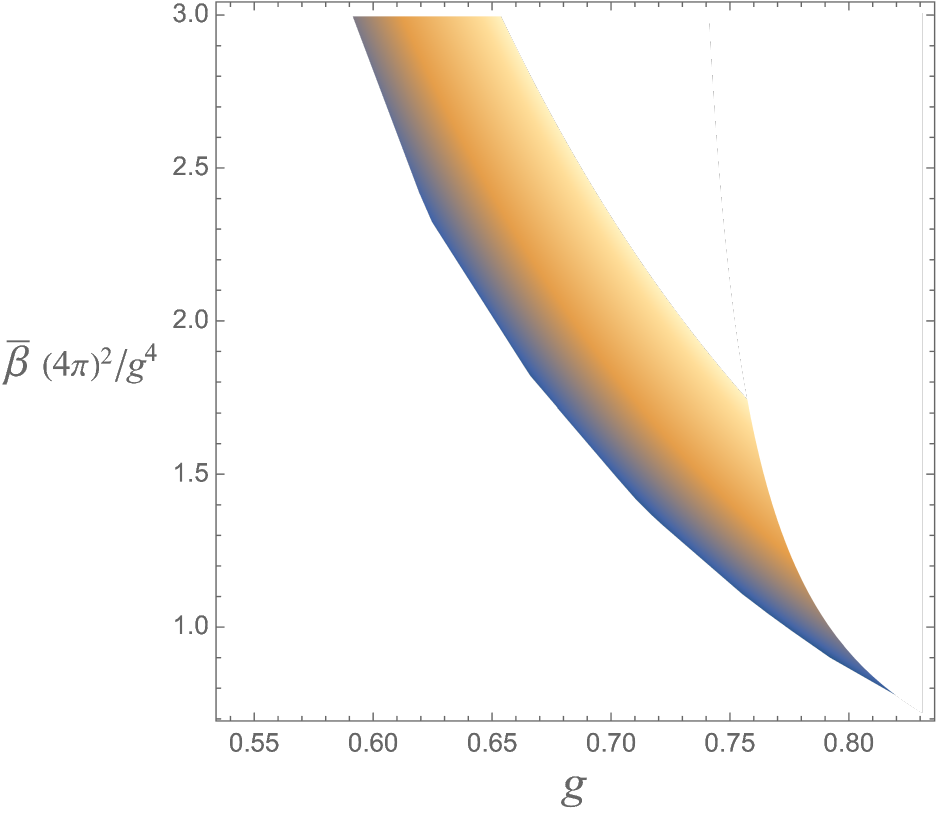}\includegraphics[scale=0.40]{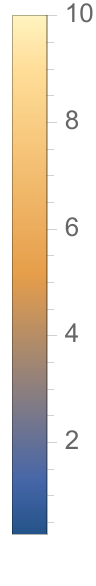}

\includegraphics[scale=0.47]{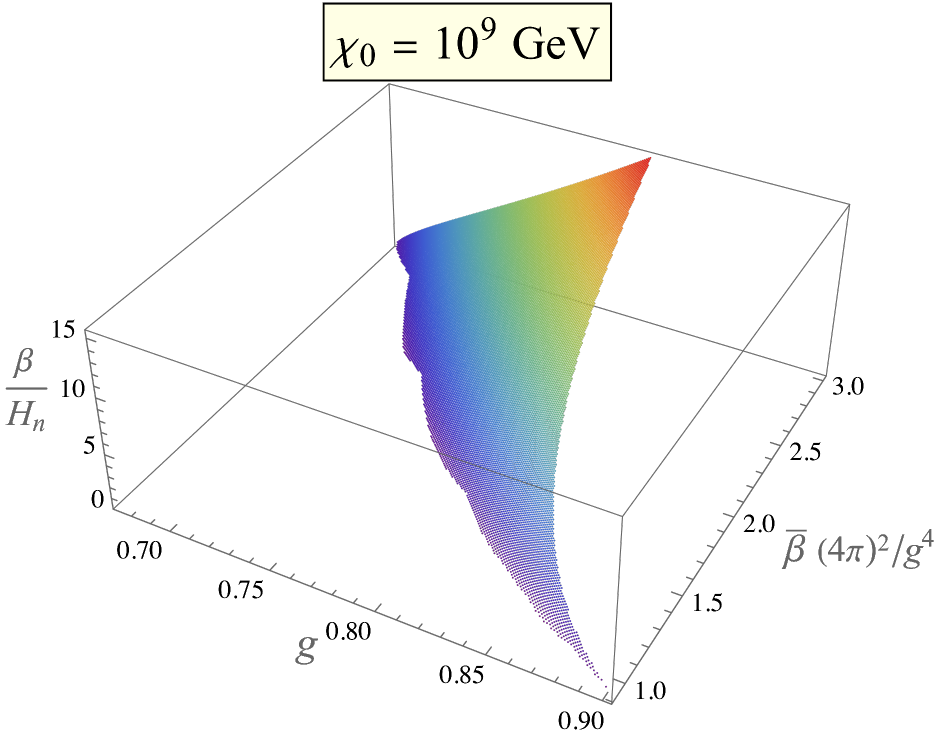}  \hspace{0.3cm}
 \includegraphics[scale=0.38]{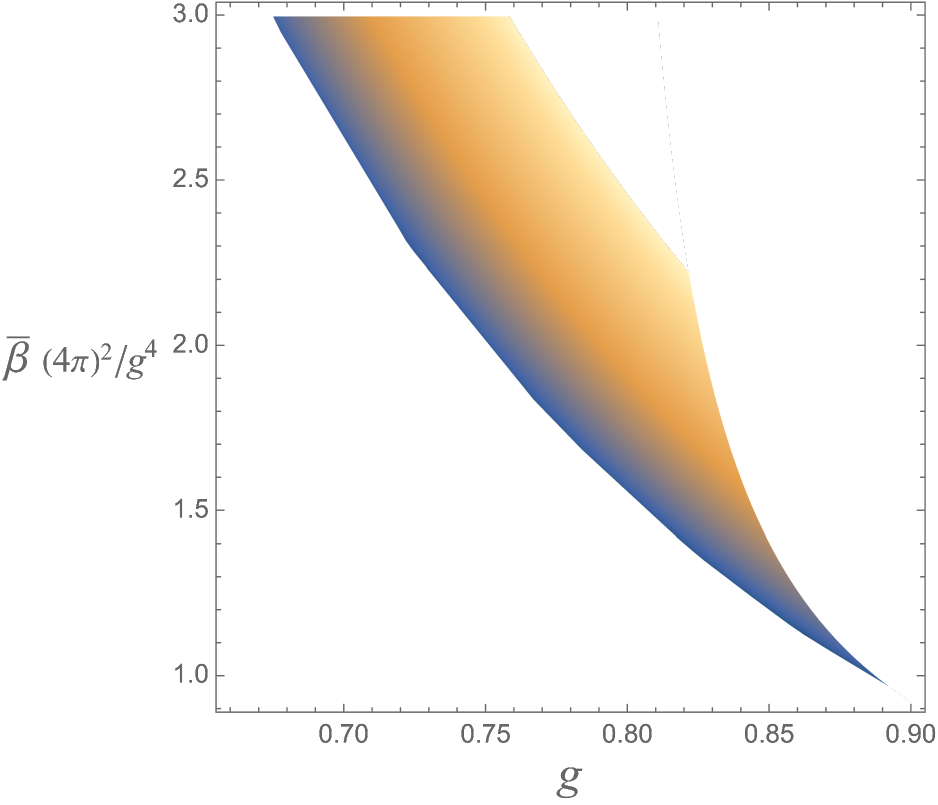}\includegraphics[scale=0.40]{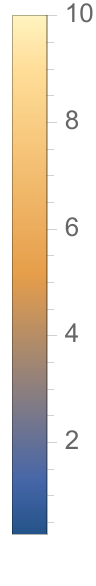}

\includegraphics[scale=0.47]{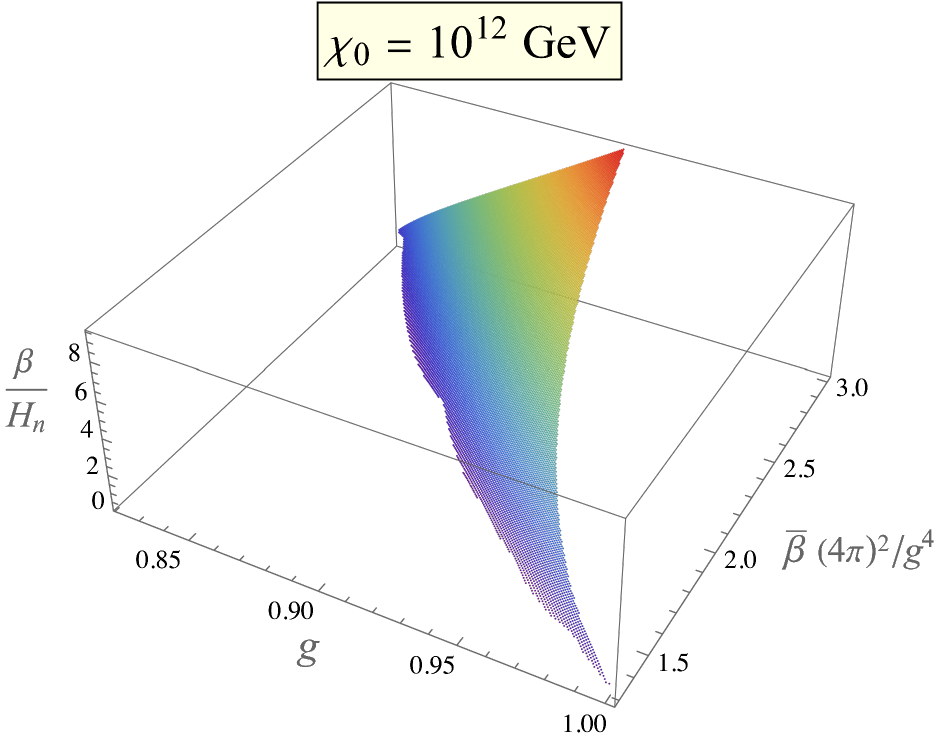}  \hspace{0.3cm}
 \includegraphics[scale=0.38]{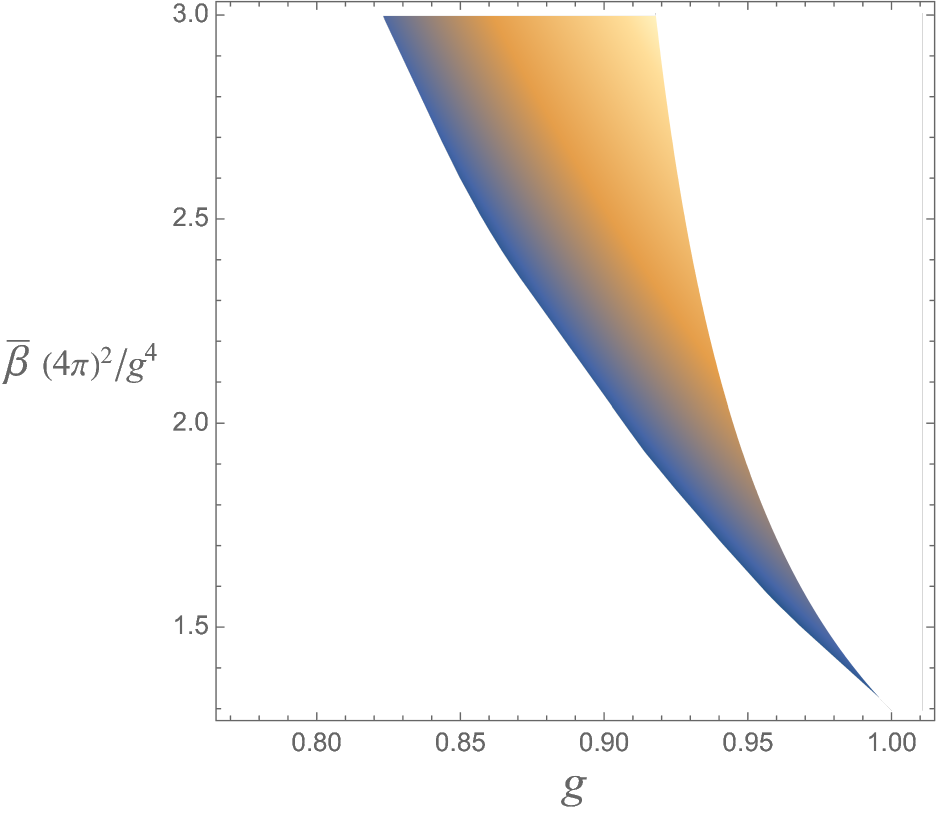}\includegraphics[scale=0.40]{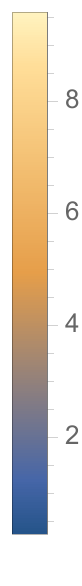}
   \vspace{-0.4cm}
    \caption{\em Like in Fig.~\ref{betagb}, but for larger values of the symmetry breaking scale $\chi_0$.}
    \label{betagbp}
  \end{center}
\end{figure}

 Figs.~\ref{betagb} and~\ref{betagbp} show $\beta/H_n$ (computed with the improved supercool expansion) as a function of $g$ and $\bar\beta$ for fixed values of $\chi_0$. There $\tilde g$ has been set equal to $g$: when $\tilde g$ is significantly lower than $g$ the expansion developed in~\cite{Salvio:2023qgb} works well as discussed in Sec.~\ref{Several degrees of freedom} and there is no need to resort to the improved supercool expansion. Moreover, in  Figs.~\ref{betagb} and~\ref{betagbp} only values of $g$ and $\bar\beta$ with $\epsilon<3$ are displayed\footnote{In the present improved approximation $\epsilon$ is computed by using Eq.~(\ref{lambda-eps}) with $\tilde \lambda = \tilde \lambda_n$}: indeed, for large values of $\epsilon$ one needs to take into account the higher-order corrections for a good accuracy. In Figs.~\ref{betagb} and~\ref{betagbp} $\beta/H_n$ never vanishes although there are values of $g$ and $\bar\beta$ for which $\beta/H_n\sim 1$. The relevant solution of the nucleation temperature equation in~(\ref{lteq}), i.e.  $\tilde\lambda_n$, ceases to exist before $\beta/H_n$ vanishes. As commented in the last paragraph of Sec.~\ref{Tnimpro}, when the solution $\tilde \lambda_n$ does not exist the effect of the spacetime curvature, as well as quantum fluctuations, can eventually become important in the decay rate.
 
 \section{Applications}\label{Applications}
 
Let us now apply the improved approximations developed in Sec.~\ref{impro} to the production of GWs and PBHs. 
 \subsection{Gravitational Waves}\label{Gravitational Waves}

   In the RSB scenario the dominant source of GWs are vacuum bubble collisions: the energy density of the space 
  where the bubbles move is dominated by the vacuum energy density due $\chi$, which leads to an exponential growth of the corresponding cosmological scale factor. This inflationary behavior dilutes preexisting matter and radiation and, thus, we neglect the GW production due to turbulence and sound waves in the cosmic fluid~\cite{Caprini:2015zlo,Maggiore:2018sht,Lewicki:2022pdb}.

In the presence of supercooling and for $\alpha\gg 1$ one finds the following GW spectrum 
due to vacuum bubble collisions\footnote{The spectral density $\Omega_{\rm GW}$ is defined as usual as
\be \Omega_{\rm GW}(f) \equiv \frac{f}{\rho_{\rm cr}}\frac{d\rho_{\rm GW}}{df},\ee 
where $\rho_{\rm cr}\equiv	3H_0^2\bp^2$  is the critical energy density, $H_0$ is the present value of the Hubble rate and $\rho_{\rm GW}$ is the energy density of the stochastic background.}~\cite{Caprini:2015zlo}
\be \label{eq:gw_col} h^2 \Omega_{\rm GW}(f) \approx 1.29 
\tt 10^{-6}\left(\frac{H_r}{\beta}\right)^2\left(\frac{100}{g_*(T_r)}\right)^{1/3}\frac{3.8(f/f_{\rm peak})^{2.8}}{1+2.8(f/f_{\rm peak})^{3.8}},\ee 
where $T_r$ is the reheating temperature after supercooling, $H_r$ is the corresponding Hubble rate and $f_{\rm peak}$ is the red-shifted frequency peak today,  given by~\cite{Caprini:2015zlo} 
\be f_{\rm peak} \approx 3.79\, \frac{\beta}{H_r}\left( \frac{g_*(T_r)}{100}\right)^{1/6}\frac{T_r}{10^{8}{\rm GeV}}  \, {\rm Hz} . \ee
Ref.~\cite{Caprini:2015zlo} used  the results of~\cite{Huber:2008hg}  based on the envelope approximation. This is an  approximation where all the energy is assumed to be stored in the bubble walls, that are taken to be thin, and at bubble collision one uses as a source for GW production the energy-momentum tensor of the uncollided part of the bubble walls. Studying the collision of two bubbles in a scalar field model with symmetry breaking entirely due to the standard Higgs mechanism, Ref.~\cite{Kosowsky:1991ua,Kosowsky:1992rz} found that  this has about 5\% accuracy. For $\epsilon\sim1$ this is comparable with the precision of the improved supercool expansion  when one  uses the approximation in~(\ref{logApp})  and neglects the terms in  the small-$x$ expansions of~(\ref{JBdef}) and~(\ref{JFdef}) of order higher than $O(x^{3/2})$. 
In our situation the envelope approximation is expected to capture the dominant source\footnote{However, see the recent works~\cite{Jinno:2017fby,Konstandin:2017sat,Lewicki:2019gmv,Lewicki:2020jiv,Lewicki:2020azd} that improved the calculation of $\Omega_{\rm GW}$ and can be relevant in the general case.}  of GWs~\cite{Freese:2022qrl} because, during the exponential growth of the universe, the bubbles expand considerably and in this process the energy gained in the transition  from the false to the true vacuum is transferred to the bubble walls, which, at the same time, become thinner for energy reasons.

For sufficiently fast reheating
\be H_r\approx H_n \approx H_I, \quad \mbox{and} \qquad T_r^4 \approx \frac{15 \bar\beta \chi_0^4}{8\pi^2 g_*(T_r)}.\label{TRHmax}\ee But otherwise $H_r$ and $T_r$ can depend on the details of the  specific model. Reheating can occur e.g. thanks to the Higgs portal coupling in~(\ref{portal}) or other portal interactions such as a kinetic mixing between the photon and a  dark photon (see~\cite{Fabbrichesi:2020wbt} for a review) that become massive through RSB. Note also that the dependence  of $\Omega_{\rm GW}$ and $f_{\rm peak}$ on $g_*(T_r)$ is quite weak.
  
  Ref.~\cite{Salvio:2023qgb} computed $f_{\rm peak}$ and provided regions where $\Omega_{\rm GW}(f_{\rm peak})$ is above the sensitivities of several current and proposed GW detectors (including LIGO, Virgo, LISA, ET, CE, BBO and DECIGO); moreover, Ref.~\cite{Salvio:2023qgb} found corresponding regions in the space of $g$, $\bar\beta$, $\chi_0$ and $\tilde g$ using the supercool expansion at LO and NLO. Then here we focus on the improved supercool expansion.

\begin{figure}[t!]
\begin{center}
 \vspace{1cm}
  \includegraphics[scale=0.52]{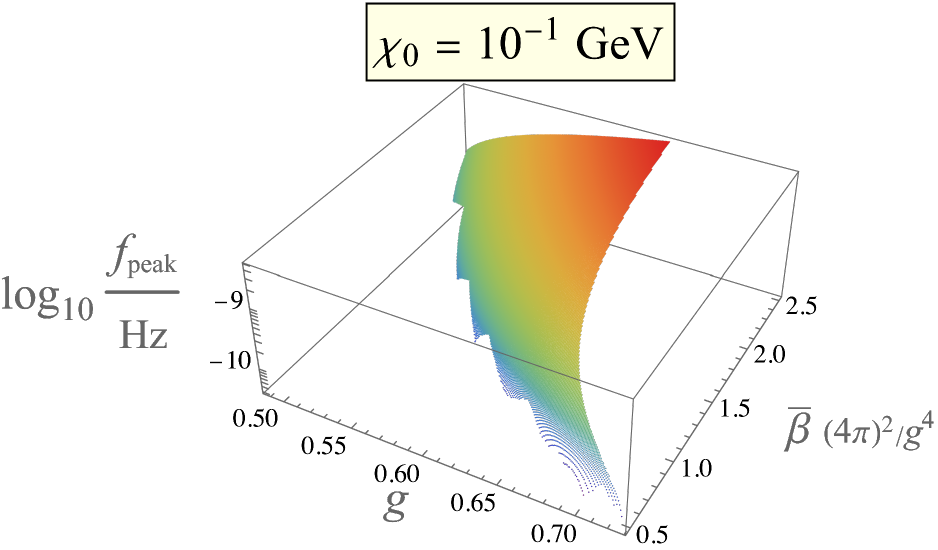}  \hspace{0.3cm}
  \includegraphics[scale=0.52]{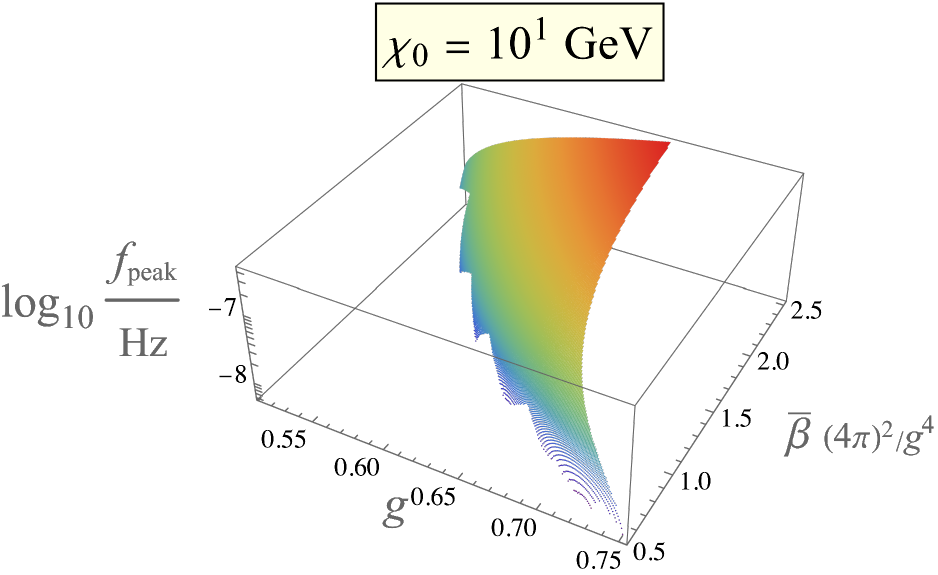} \\ \includegraphics[scale=0.52]{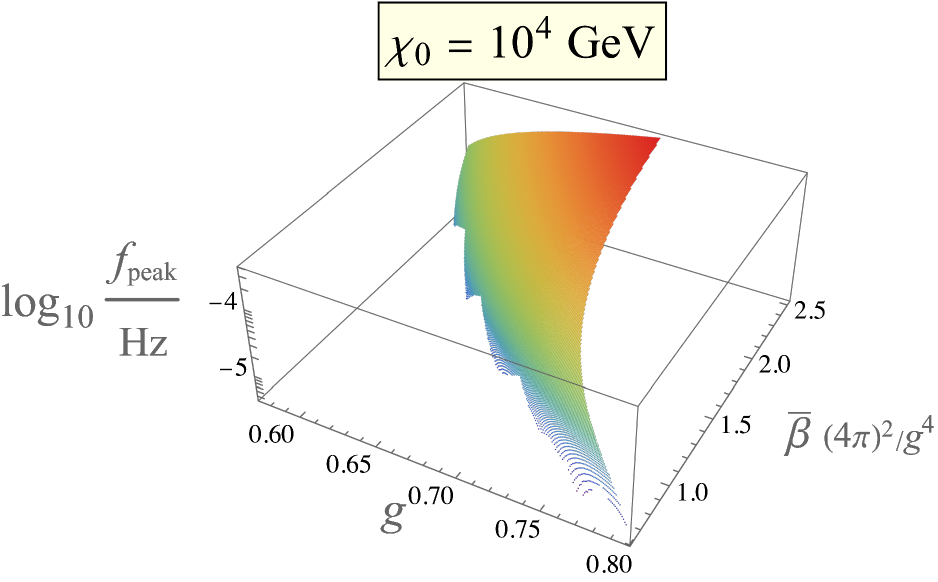}  \hspace{0.3cm}
  \includegraphics[scale=0.54]{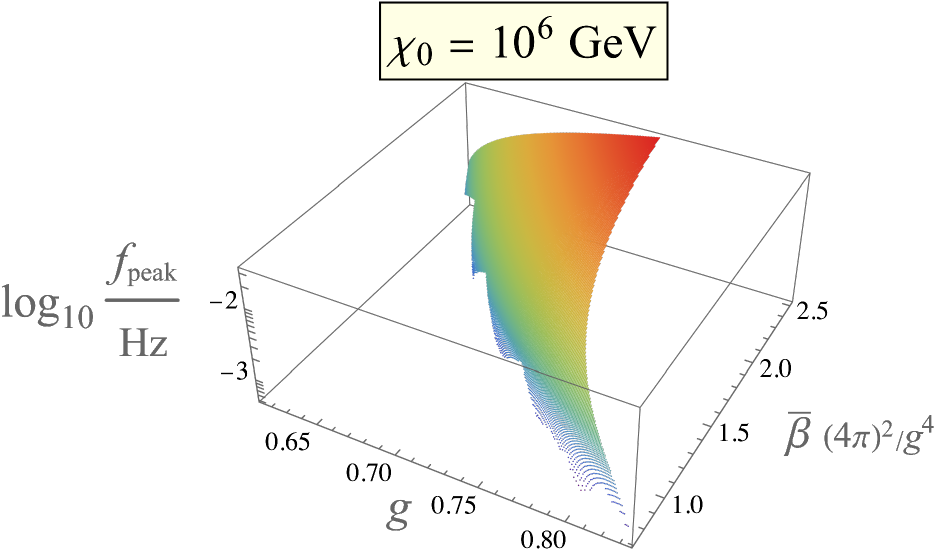} \\
  \includegraphics[scale=0.52]{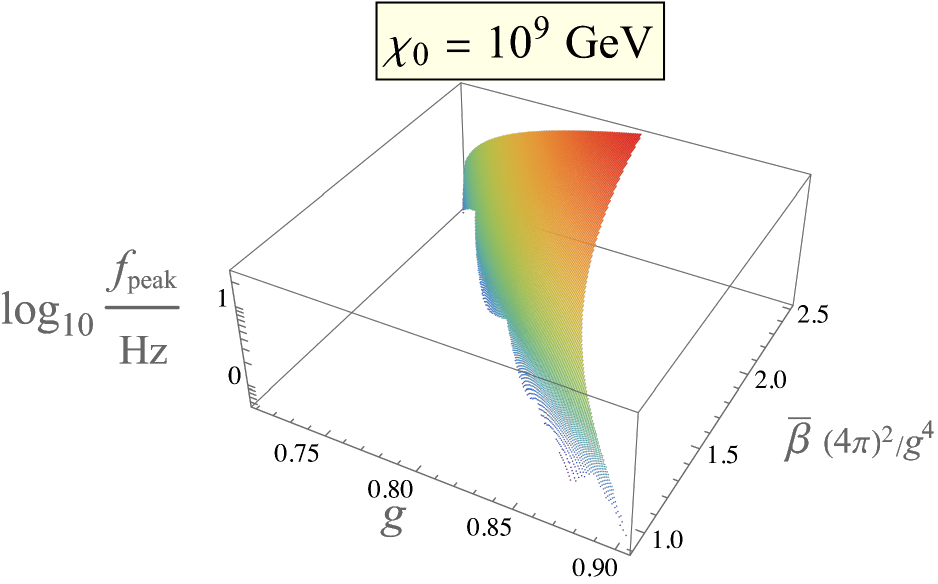}  \hspace{0.3cm}
  \includegraphics[scale=0.54]{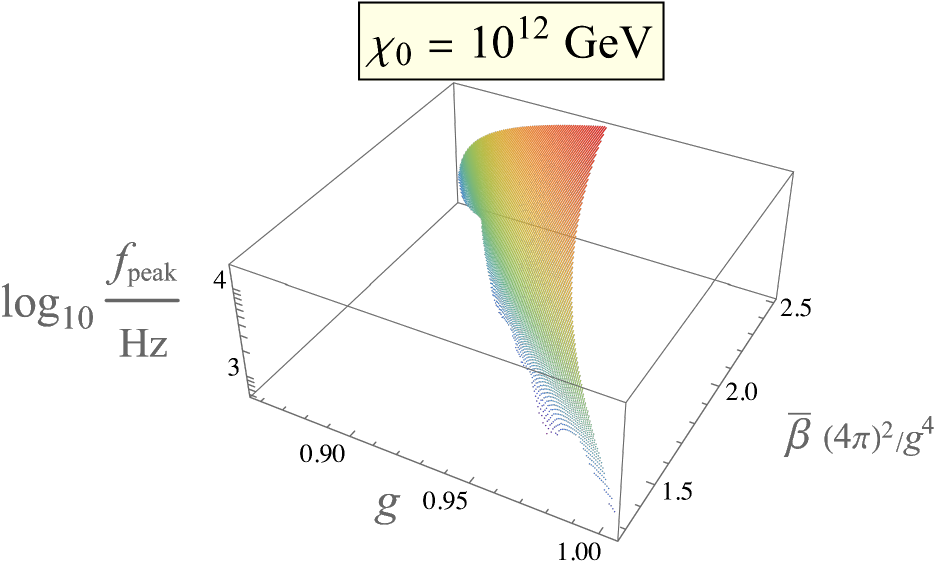}   
      \caption{\em The peak frequency as a function of $g$ and $\bar\beta$ in the case of fast reheating and fixing $g_*(T_r)=110$.  Also, $\tilde g = g$ and $\epsilon<3$ has been imposed.}\label{fpimpro}
  \end{center}
\end{figure}

 In Fig.~\ref{fpimpro} $f_{\rm peak}$ computed with the improved supercool approximation is shown  for various values of $\chi_0$; moreover, in that figure we considered only values of $g$ and $\bar\beta$ such that $\epsilon<3$  and set $\tilde g = g$ (for the reasons explained at the end of Sec.~\ref{Duration of the phase transition}).  Fig.~\ref{fpimpro} also shows frequencies of GW signals that have been recently detected by pulsar timing arrays~\cite{NANOGrav:2023gor,Antoniadis:2023ott,Reardon:2023gzh,Xu:2023wog} (see Ref.~\cite{NANOGrav:2023hvm} for a PT interpretation of the detected signals performed by the NANOgrav collaboration and relevant for our study and Refs.~\cite{Antoniadis:2023zhi,Bringmann:2023opz,Madge:2023cak,Zu:2023olm,Han:2023olf,Fujikura:2023lkn,Kitajima:2023cek,Bai:2023cqj,Addazi:2023jvg,Athron:2023mer,Lu:2023mcz,Xiao:2023dbb,Li:2023bxy,Ghosh:2023aum,Wu:2023hsa,DiBari:2023upq} for other independent discussions of PT interpretations).

\begin{figure}[t!]
\begin{center}
      \includegraphics[scale=0.51]{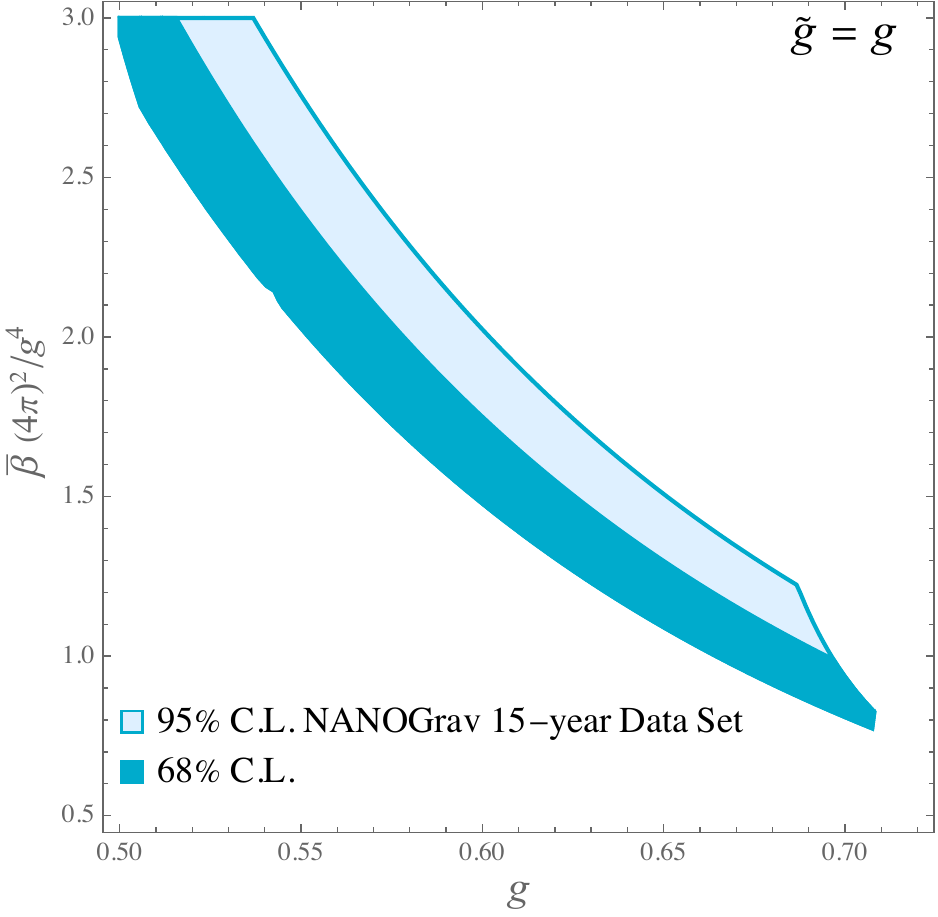}  \hspace{1cm}
  \includegraphics[scale=0.51]{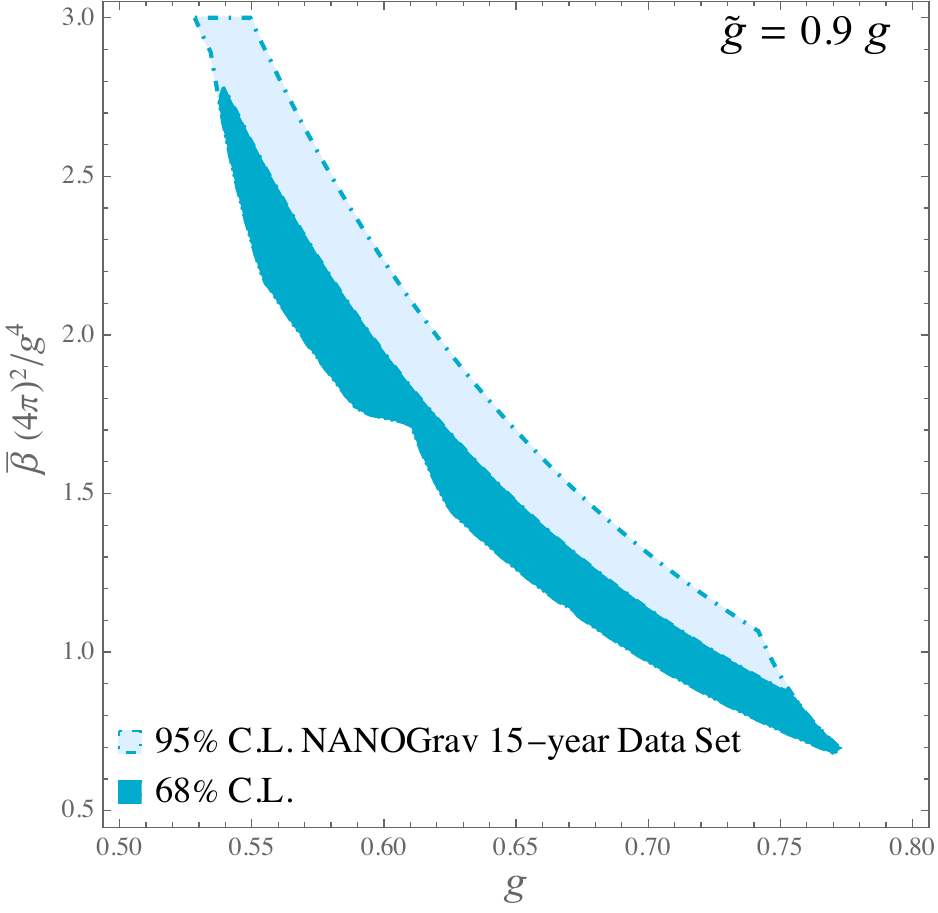} 
      \caption{\em Regions corresponding to the GW background detected by pulsar timing arrays. In both plots $\chi_0=10$~GeV, $g_*(T_r)=110$ and fast reheating is assumed. Here  $\epsilon<3$ has been imposed.}\label{NANO-impro}
  \end{center}
\end{figure}

 Combining with the information in Fig.~\ref{betagb}, one finds that it is possible to account for the signals  detected by pulsar timing arrays for $\chi_0\sim 10$~GeV and choosing the basic parameters  $g$, $\tilde g$ and $\bar\beta$ appropriately, as illustrated in Fig.~\ref{NANO-impro}.

\begin{figure}[t!]
\begin{center}
      \includegraphics[scale=0.5]{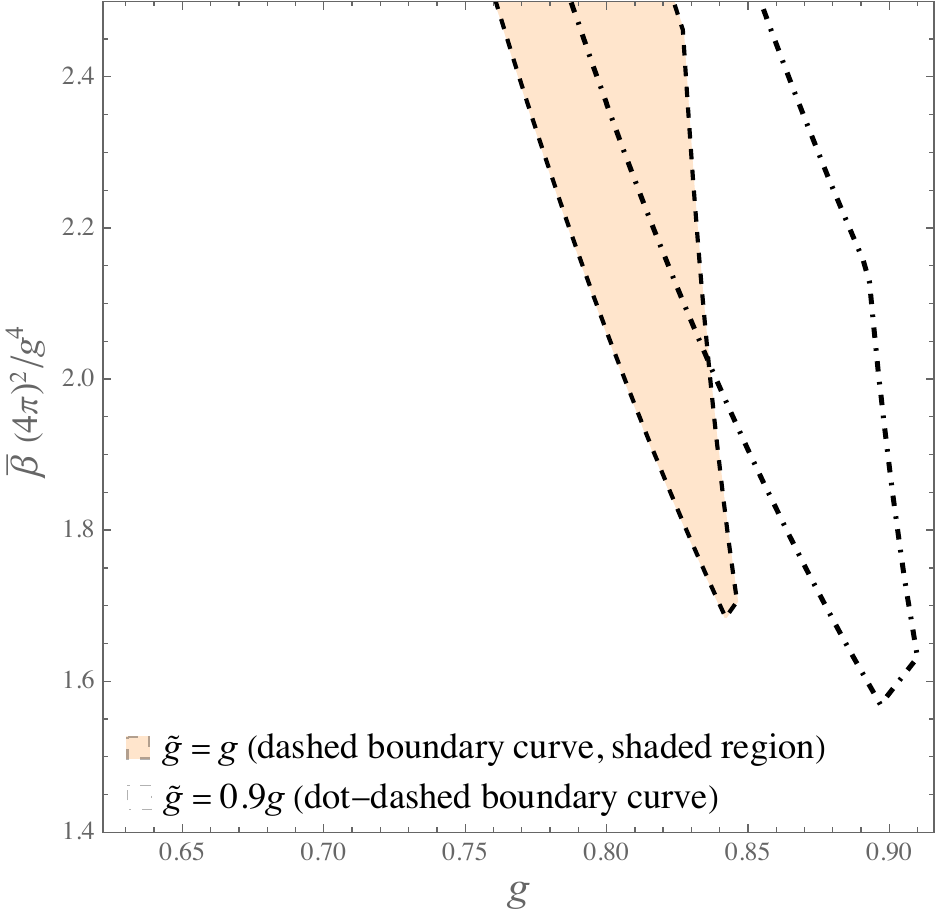}  \hspace{1cm}
  \includegraphics[scale=0.5]{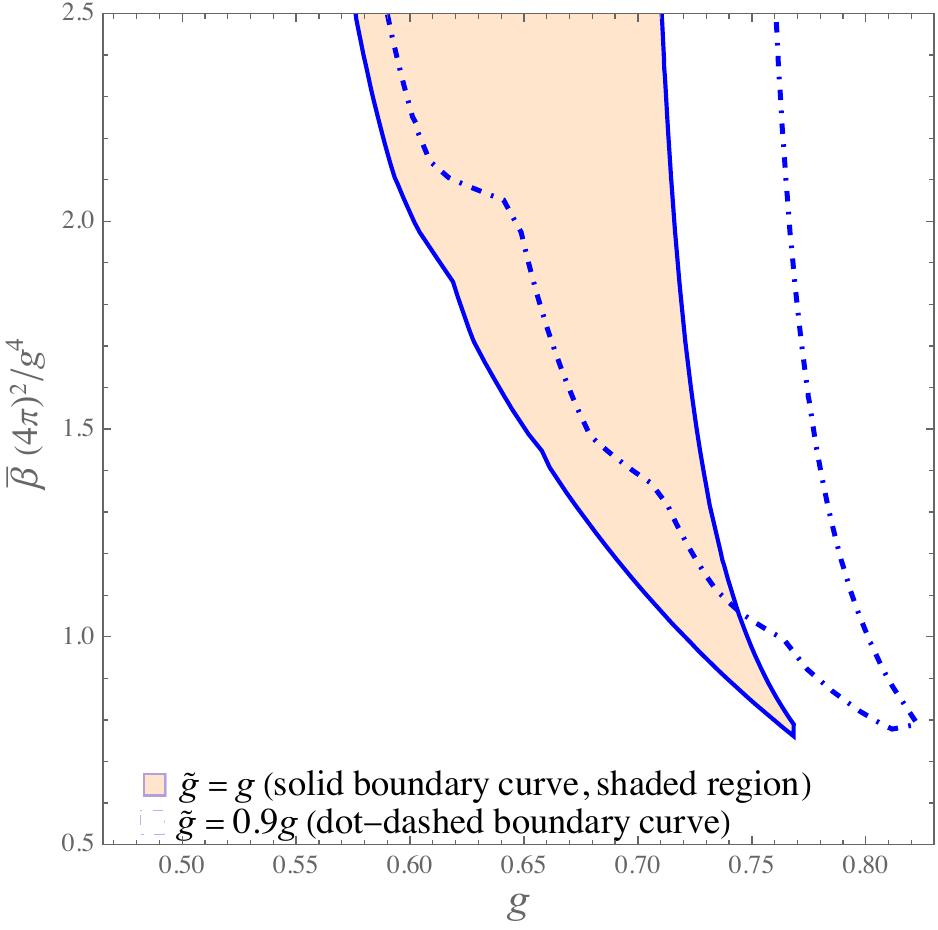} 
      \caption{\em Regions where $\Omega_{\rm GW}(f_{\rm peak})$ is above the sensitivities of LIGO-VIRGO O3 (left plot, where $\chi_0 = 2 \times 10^9$~GeV) and LISA (right plot, where $\chi_0 = 10^4$~GeV). In both plots $g_*(T_r)=110$ and fast reheating is assumed. Here  $\epsilon<3$ has been imposed.}\label{LIGO-LISA-impro}
  \end{center}
\end{figure}

Fig.~\ref{LIGO-LISA-impro} shows instead the regions where $\Omega_{\rm GW}(f_{\rm peak})$ is above the sensitivities of Advanced LIGO's and Advanced Virgo's third observing (O3) run (left plot) and LISA with power law sensitivity~\cite{Babak:2021mhe} (right plot) for two non-vanishing values of $\tilde g$ and using the improved supercool approximation. The regions in the left plot are thus, remarkably, ruled out.  In Fig.~\ref{LIGO-LISA-impro} we again considered only values of $g$ and $\bar\beta$ such that $\epsilon<3$. The parameter $\chi_0$ has been chosen around $10^9$~GeV in the left plot of Fig.~\ref{LIGO-LISA-impro} and $10^4$~GeV in the right one because the corresponding $f_{\rm peak}$ is then around the frequency range of LIGO-VIRGO O3~\cite{KAGRA:2021kbb} and LISA~\cite{Babak:2021mhe}, respectively (see Fig.~\ref{fpimpro}). A $\chi_0$ around  $10^9$~GeV is relevant e.g.~for axion models, while a $\chi_0$ around 10 or 100 TeV could be associated with observable physics at colliders and  is relevant e.g. for supersymmetric models or low-scale unified theories such as the Pati-Salam model~\cite{Pati:1974yy}
  or Trinification~\cite{Babu:1985gi}.

\subsection{Primordial black holes}\label{Primordial black holes}

As shown in~\cite{Salvio:2023qgb} the PT associated with an RSB is always of first order. Besides having the potential of leaving observable GW footprints, first-order PTs can also naturally generate PBHs because generically lead to large over-densities~\cite{Hawking:1982ga,Crawford:1982yz,Kodama:1982sf,Moss:1994iq,Gross:2021qgx,Kawana:2021tde,Liu:2021svg,Baker:2021sno,Jung:2021mku,Hashino:2021qoq,Kawana:2022olo,Gouttenoire:2023naa}. One of the main motivations for studying PBHs is the fact that they can account for a fraction $f_{\rm {PBH}}$ of (or even the whole) dark matter density.

\subsubsection{Late-blooming mechanism}
One of the mechanism to generate PBHs from first-order PTs is based on the presence of strong supercooling, which generically takes place in the RSB scenario and is a key property for the validity of the supercool expansion. Since the bubble formation process is statistical for both quantum and thermal reasons, distinct causal patches percolate at different times. Patches that percolate the latest undergo the longest vacuum-dominated stage and, therefore, develop large over-densities triggering their collapse into PBHs. This late-blooming mechanism has been studied in a number of papers (see e.g. Refs.~\cite{Kodama:1982sf,Liu:2021svg,Hashino:2021qoq,Kawana:2022olo,Gouttenoire:2023naa}) and we refer the reader to these works for an introduction to this mechanism. A key feature is that the longer the supercooling period lasts (the smaller $\beta/H_n$ is) the more effective this mechanism is.

\begin{figure}[t!]
\begin{center}
   \hspace{-0.5cm}   \includegraphics[scale=0.5]{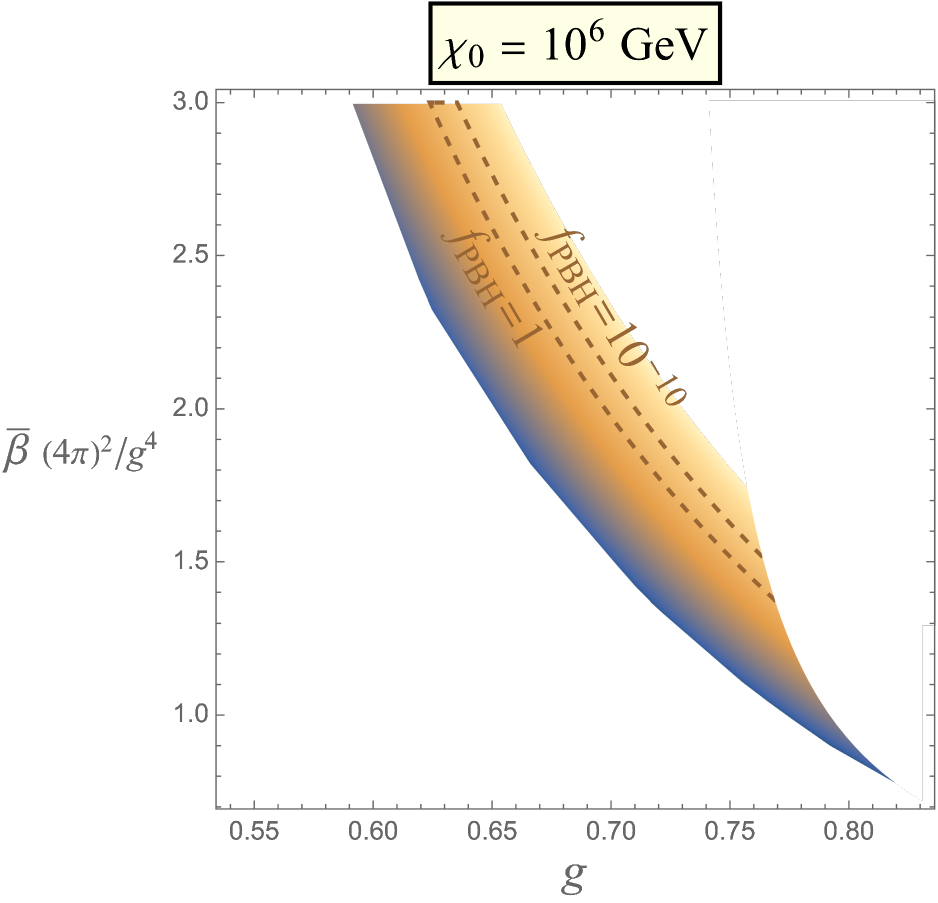}  \includegraphics[scale=0.42]{betagb106L.pdf}\hspace{0.5cm}
  \includegraphics[scale=0.5]{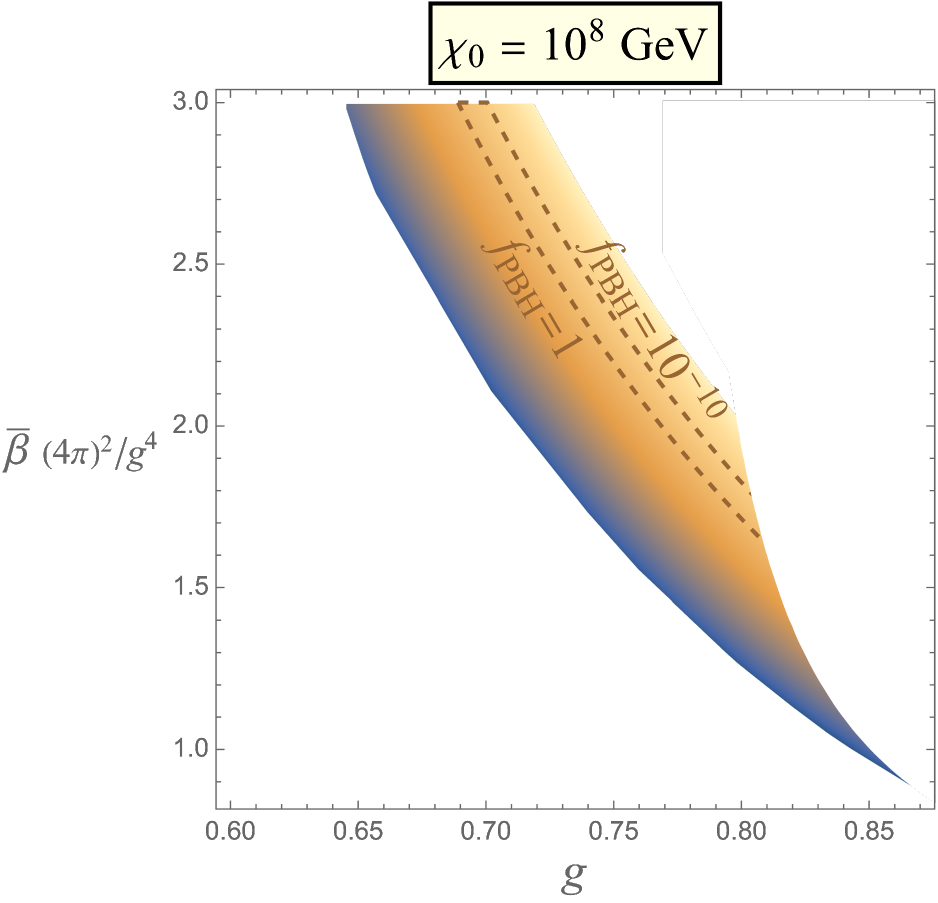} 
  \includegraphics[scale=0.42]{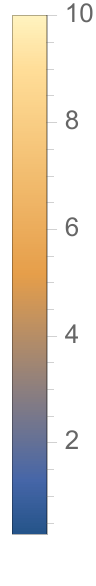}
      \caption{\em Density plots giving the values of $\beta/H_n$ varying $g$ and $\bar\beta$. On the lower dashed line the whole dark matter is due to PBHs generated through the late-blooming mechanism ($f_{\rm PBH}=1$); the upper dashed line corresponds instead to $f_{\rm PBH}=10^{-10}$. Here $\tilde g=g$ and $\epsilon<3$ has been imposed.}\label{PBH}
  \end{center}
\end{figure}

Following the method illustrated in Ref.~\cite{Gouttenoire:2023naa} and using the improved supercool expansion  we have identified regions of the parameter space (shown in Fig.~\ref{PBH}) for which PBHs produced by the late-blooming mechanism can account for a significant fraction of the dark matter density in a model-independent way. This was possible because, as discussed in Sec.~\ref{Duration of the phase transition} and shown in Figs.~\ref{betagb} and~\ref{betagbp}, we can compute $\beta/H_n$ only in terms of the parameters $g$, $\tilde g$, $\bar\beta$ and $\chi_0$ for large-enough supercooling: this hypothesis allows us to use the supercool expansion (in Fig.~\ref{PBH}  the improved version is used).  For all values of these parameters in Fig.~\ref{PBH} the PT is very strong ($\alpha>100$ for $g_*\sim10^2$) and the improved supercool expansion gives a good approximation for the key quantities of the PT in a model-independent way. The regions of Fig.~\ref{PBH} contained between the dashed lines have $10^{-10}< f_{\rm PBH}<1$. The regions below the lower dashed line, for which $f_{\rm PBH}=1$, are, remarkably, excluded in a model independent way because of the phenomenological necessity of not overproducing dark matter.

\subsubsection{Other mechanisms?}
Several other mechanisms to produce PBHs have been proposed in the literature. Some of these are unrelated to the RSB and strong supercooled PTs and thus we do not discuss them, although, of course, they could contribute to the PBH abundance in specific models. 

Another mechanism that can be a priori related to the RSB  and strong supercooled PTs in a model-independent way is the one based on bubble collisions~\cite{Hawking:1982ga,Moss:1994iq,Jung:2021mku}. However, in Ref.~\cite{Jung:2021mku} it was pointed out that bubble collisions during a first-order PT can produce PBHs only if the bubble radii become near-horizon-sized and the bubble walls have a non-negligible thickness when they collide. In RSB PTs this PBH production mechanism is, therefore, suppressed because the bubble walls become very thin after a long period of supercooling (as discussed in Sec.~\ref{Gravitational Waves})  and also we checked that the bubble radii never become near-horizon-sized for values of $\chi_0$ up to $10^{16}$~GeV. Larger values  of $\chi_0$ are not considered here as they require a UV completion of gravity.

\section{Examples of specific models}\label{Examples of specific models}

What we have done so far is a model-independent study of phase transitions and corresponding production of GWs and PBHs in the RSB scenario, which is valid in the supercool expansion or, more generally, in the improved supercool expansion.
  This formalism can be applied to any RSB model featuring a large-enough amount of supercooling ($\epsilon$ at most of order 1).
  To illustrate the usefulness of these results here we apply them to some concrete models. 

\subsection{A simple model}\label{A simple toy model}

We start with a simple toy model that can illustrate all essential features of the RSB scenario. The basic requirements of RSB is the existence of a flat direction that is radiatively broken to generate a minimum ($\bar \beta >0$). This positivity condition can be satisfied by introducing a gauge group, which can generically give positive contributions to the scalar beta functions. Here we take this group to be $SU(2)$ (an Abelian case will be discussed in Sec.~\ref{Radiative electroweak and lepton symmetry breaking}). The scalar fields will be organized here in a complex adjoint field $A$, whose no-scale potential is\footnote{A global U(1) symmetry acting on $A$ is imposed to forbid additional terms.} 
\be V_{\rm ns} = \lambda_1 {\rm Tr}^2(A^\dagger A)  + \lambda_2|{\rm Tr}(A A)|^2,\ee 
where $\lambda_1$ and $\lambda_2$ are real couplings.
Therefore, there exist a non-trivial flat direction, for which $A=A^\dagger$, at a scale $\tilde \mu$ where $\lambda_1+\lambda_2=0$. When $A=A^\dagger$ the three components $A_k$ of $A$ along the Pauli matrices, $A=A_k\sigma^k/2$, can always be transformed through an element of the gauge group SU(2) in a way that only one of these components is not vanishing and positive. We identify this non-zero component with $\chi$.

Here $\bar\beta$ is the beta function of $\lambda_1+\lambda_2$, i.e.
\be \mu\frac{d}{d\mu}(\lambda_1 +\lambda_2) = \frac1{(4\pi)^2}\left[12 g_a^4+40\lambda_1\lambda_2-24g_a^2(\lambda_1+\lambda_2)+28(\lambda_1^2+\lambda_2^2)\right],\ee 
where $g_a$ is the gauge coupling of SU(2).
Evaluating at the scale $\tilde \mu$, at which $\lambda_2=-\lambda_1$, gives
\be \bar\beta =  \frac1{(4\pi)^2}\left[12 g_a^4+16\lambda_1^2\right]_{\mu=\tilde\mu}. \ee
In order to simplify the following discussion we also assume that $\lambda_1\ll g_a$ such that we have a single coupling to deal with. 

In this case the massive background-dependent spectrum only features two spin-1 particles with equal mass, $M_V =|g_a| \chi$. All the other masses either vanish or are negligibly smaller. So the collective coupling $g$ defined in Eq.~(\ref{M2g2def}) turns out to be 
\be g= \sqrt{6} |g_a| \ee
and so
\be \bar\beta = \frac{g^4}{3(4\pi)^2}. \ee
Also, $\tilde g$ defined in Eq.~(\ref{gtdef}) reads 
\be \tilde g = \sqrt[3]{6} |g_a|=\frac{g}{\sqrt[6]{6}}. \ee 
Having determined $\bar\beta$ and $\tilde g$ in terms of $g$ one can now use the model-independent analysis based on the improved supercool expansion of Sec.~\ref{Improved supercool expansion} with only two free parameters: $g$ and $\chi_0$.

\begin{figure}[t!]
\begin{center}  
\includegraphics[scale=0.45]{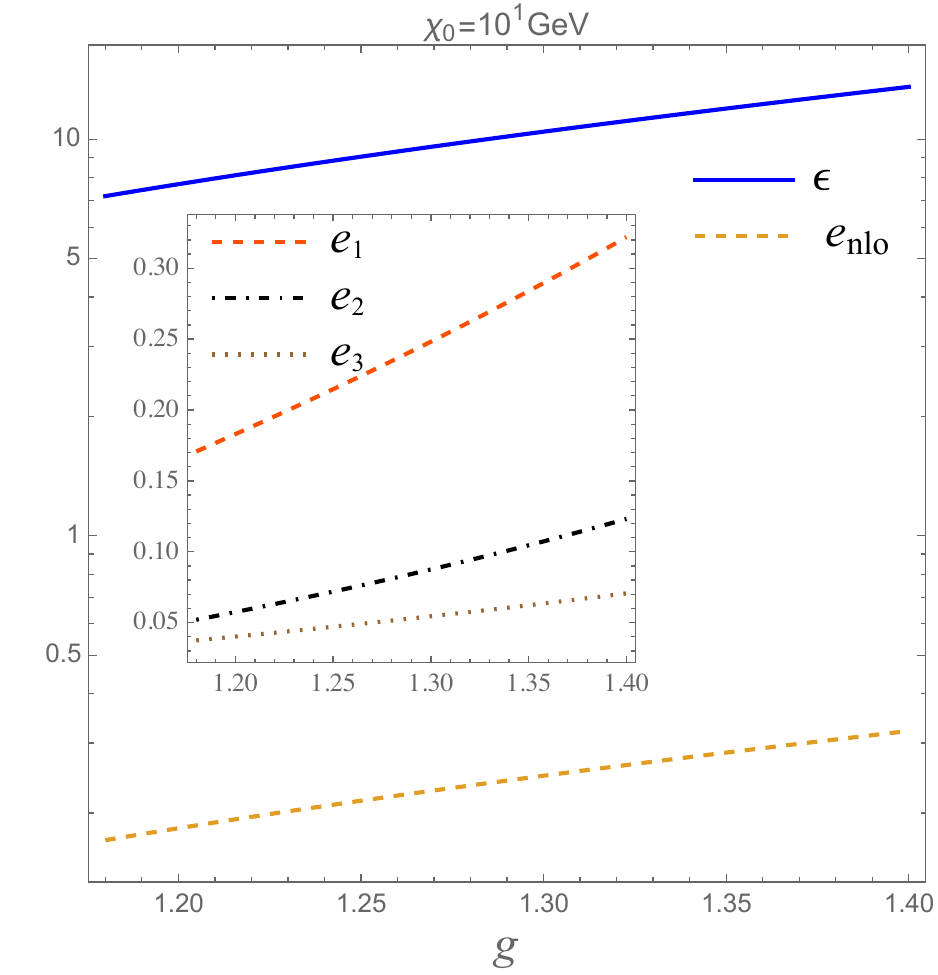}  
\hspace{1cm}
  \includegraphics[scale=0.45]{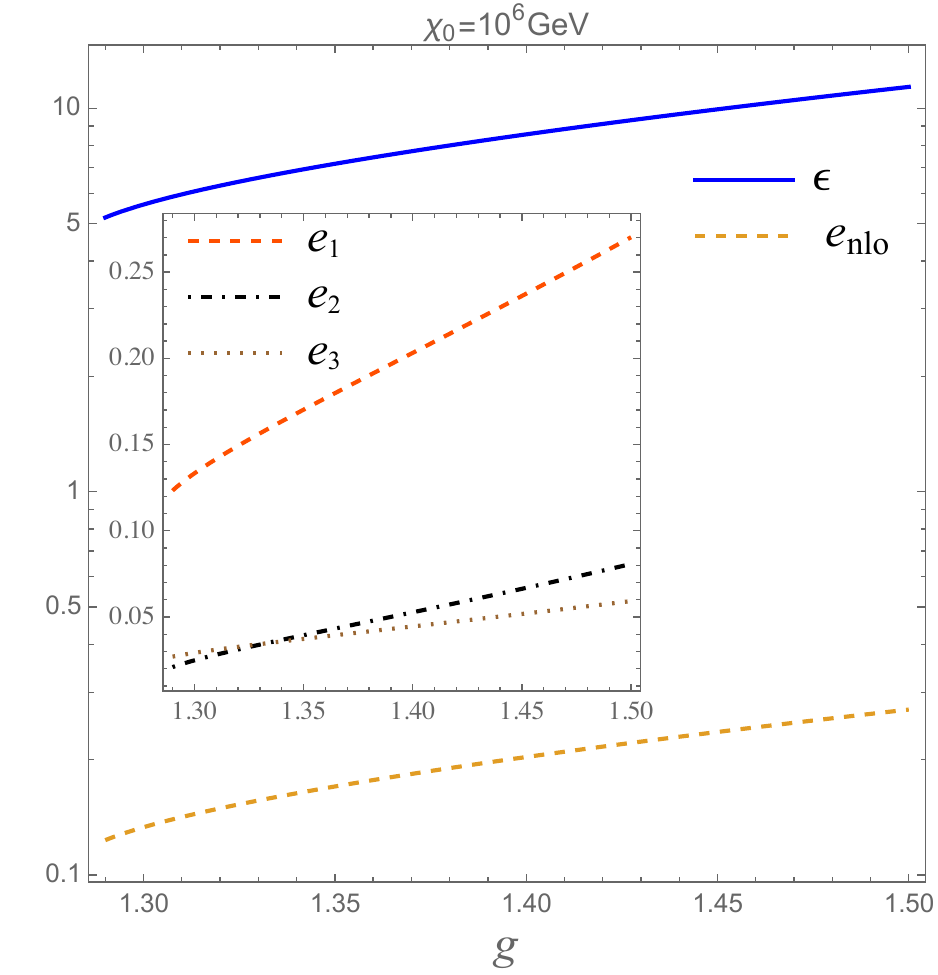} \\
   \includegraphics[scale=0.45]{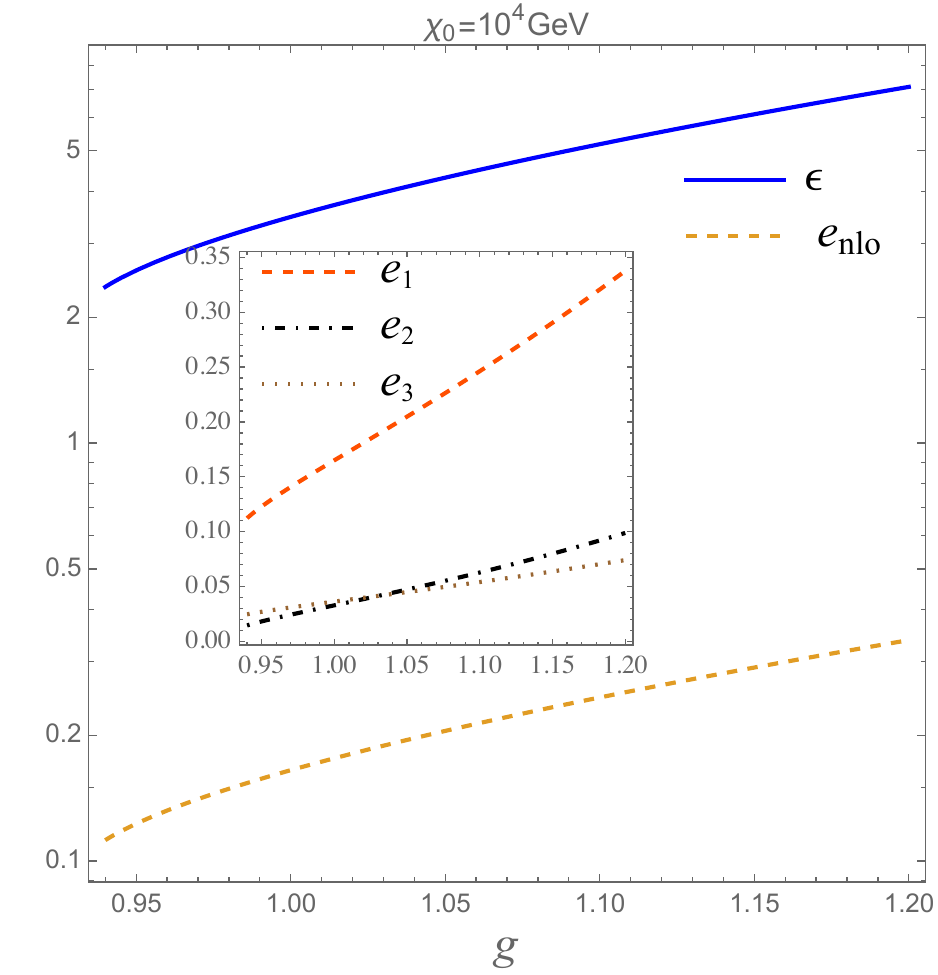}  \hspace{1cm} \includegraphics[scale=0.45]{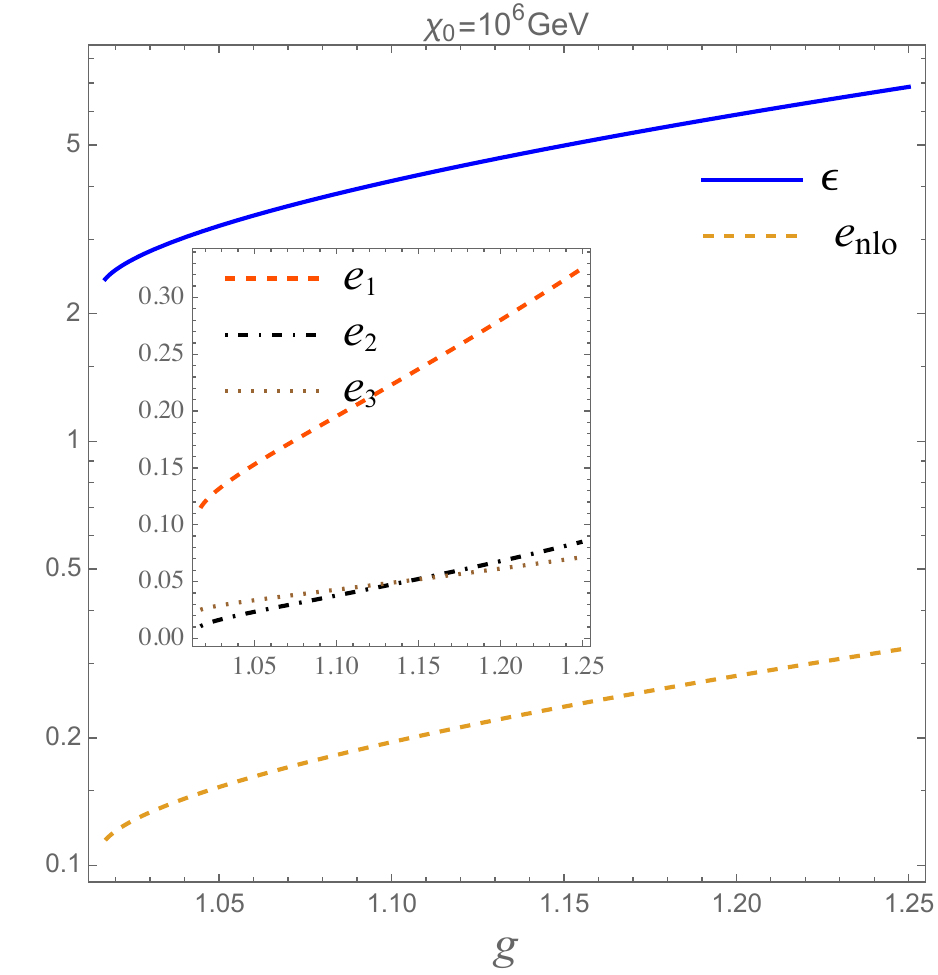}  
    \caption{\em Comparison between $\epsilon$ and the error $e_{nlo}$  that one is making in using the standard supercool expansion at NLO of Sec.~\ref{recap}.  The upper plots refer to the simple toy model of Sec.~\ref{A simple toy model} ($e_{nlo}$ is defined in~(\ref{enlos})), while the lower plots regards Sec.~\ref{Radiative electroweak and lepton symmetry breaking} ($e_{nlo}$ is defined in~(\ref{enlo})).}\label{enloBL}
  \end{center}
\end{figure}

At this point it is interesting to quantify the error that one is making in analysing this model with the standard supercool expansion at NLO of Sec.~\ref{recap} rather than with the improved supercool expansion of Sec.~\ref{Improved supercool expansion}, namely treating the cubic term in~(\ref{barVnlo}) perturbatively. Fig.~\ref{enloBL} (upper plots) shows $\epsilon$ and 
\be e_{nlo} \equiv \max(e_1,e_2,e_3), \qquad (e_1,e_2,e_3)\equiv \left(\frac{\tilde c_3^2 \tilde g^6 \epsilon}{2\pi^2 c_3^2 g^6}, \frac{|\frac12\log(\epsilon/g^2)-1/4|}{X},\frac{\epsilon}{6\times 32}\right) \label{enlos}\ee
(computed for simplicity with the LO formula for $T_n$ in~(\ref{appTn})). The parameter $e_{nlo}$ in~(\ref{enlos}) quantifies the above-mentioned error: the first entry $e_1$ in the $\max$ function is the square of the size of the second term in~(\ref{S3k0}) relative to the first one (which is an estimate of the next-to-next-to-leading correction in treating the cubic term in~(\ref{barVnlo}) perturbatively);  the second and third entries, $e_2$ and $e_3$, are instead estimates of the error due to the approximation in~(\ref{logApp}) and to truncating the small-$x$ expansions in~(\ref{JBdef}) and~(\ref{JFdef}) up to the $x^{3/2}$ term\footnote{The extra factor of $6$ in the denominator of the last entry comes from the fact that the $x^2$ term in the small-$x$ expansions in~(\ref{JBdef}) and~(\ref{JFdef}) features a coefficient $\sum_b n_b m_b^4$ which equals $g^4\chi^4/6$ in this case.}, respectively. As one can see, although $\epsilon$ is above 1 the quantity $e_{nlo}$ is small, especially for smaller values of $g$. The reason why this happens is because here one has two massive vector fields for a total of six  degrees of freedom and there is, therefore, an extra suppression of the neglected terms as explained in Sec.~\ref{Several degrees of freedom}. Looking at Fig.~\ref{enloBL} one also sees that $e_1$ is larger than $e_2$ and $e_3$, meaning that the improved supercool expansion is a better approximation than the standard supercool expansion at NLO in this case. 
This is not surprising because the number of degrees of freedom with dominant couplings to the flat-direction field $\chi$ is not very large (it is six) and $\epsilon$ is not smaller than one in this case.

\subsection{Radiative electroweak and lepton symmetry breaking}\label{Radiative electroweak and lepton symmetry breaking}

Let us  study now an example that is  phenomenologically well-motivated. 
The SM is a very successful model but it clearly has to be extended: neutrino oscillations, dark matter and baryon asymmetry must be accounted for in a phenomenologically complete model. One of the most economical way to achieve this goal is to add three right-handed neutrinos $N_i$ featuring Majorana masses  below the EW scale~(see e.g.~\cite{Asaka:2005pn,Canetti:2012kh}).

The corresponding Majorana mass terms can be promoted to scale-invariant Yukawa interactions
\be  \frac12 y_{ij} A N_iN_j + \mbox{h.c.} \label{ANYuk}\ee 
in $\Lag^{\rm ns}_{\rm matter}$ of Eq.~(\ref{eq:Lmatterns}) by introducing a charged scalar $A$ with a non-vanishing lepton number (here the $y_{ij}$ are the corresponding Yukawa couplings). Coupling $A$ and the $N_i$ to the classically scale-invariant part\footnote{The tachyonic mass parameter of the Higgs, which is needed to induce EW symmetry breaking, emerges radiatively as described in Eq.~(\ref{muhInd}).} of the SM through renormalizable dimension-four interactions allows us to build a classically scale-invariant model of the type described in Sec.~\ref{recap}. In order to generate the $N_i$ Majorana masses one can then try to realize an RSB of the lepton number along the field direction\footnote{With two scalar fields, $A$ and the Higgs doublet $\mathcal{H}$, one can conceive other flat directions. However, such modification of the SM should appear at a sufficiently high mass scale $\chi_0$ to fulfil the experimental bounds and, therefore, the quartic portal coupling $\lambda_{ah}$ between $|A|^2$ and $|\mathcal{H}|^2$ should be sufficiently small to respect Eq.~(\ref{muhInd}) with the measured value of $\mu_h^2$. In this limit the only viable flat direction should be along $A$ for $\lambda_a=0$.} $|A|$, which, as we have seen, requires the quartic coupling $\lambda_a$ of the field $A$ to vanish at an energy scale where its beta function is positive (see Eq.~(\ref{eq:CWgen})). However, it turns out that in this simple model the Yukawa interactions in~(\ref{ANYuk}) drives this beta function  to negative values when $\lambda_a$ and $\lambda_{ah}$ are negligibly small.

This problem can be elegantly solved by gauging the Abelian U(1) symmetry acting on $A$. As well known, in order to avoid any gauge anomalies, such new gauge symmetry must correspond to $B-L$ and so we call it $U(1)_{B-L}$. Therefore all leptons (including the $N_i$) and quarks as well as the scalar $A$ are charged under this Abelian symmetry. 
A radiatively-induced vacuum expectation value of $A$ can then generate the $N_i$ Majorana masses and induce the tachyonic Higgs mass parameter in  Eq.~(\ref{muhInd}). This classically scale-invariant model has been previously considered in Ref.~\cite{Iso:2009ss}, but without accounting for dark matter. 

The Lagrangian is given by
\bea &&\Lag^{\rm ns}_{\rm SM} + D_\mu A^\dagger D^\mu A + \bar N_j i\slashed{D}N_j -\frac14 B'_{\mu\nu}B'^{\mu\nu} \nonumber \\ &&+\left( Y_{ij} L_i\mathcal{H} N_j +\frac12 y_{ij} A N_iN_j + \mbox{h.c.}\right) -\lambda_a|A|^4+\lambda_{ah} |A|^2|\mathcal{H}|^2,
\eea 
where $\Lag^{\rm ns}_{\rm SM}$  represents the classically scale-invariant SM Lagrangian and the $L_i$ are the three families of SM lepton doublets.  
Here $D_\mu$ is the covariant derivative with respect to the full gauge group $SU(3)_C\times SU(2)_L\times U(1)_Y\times U(1)_{B-L}$, i.e. the SM group times the $B-L$ one:
\be  D_\mu = \partial_\mu +i g_3 T^\alpha G^\alpha_\mu + i g_2 T^a W^a_\mu + i g_Y {\mathcal Y} B_\mu + i \left[g_m {\mathcal Y}+ g_1' (B-L)\right] B'_\mu,\label{CovBmL}\ee 
which involve the gluons $G^\alpha_\mu$, the triplet of $W$ bosons $W^a_\mu$ as well as the gauge fields $B_\mu$ and $B_\mu'$ of $U(1)_Y$ and $U(1)_{B-L}$ (as usual $B'_{\mu\nu} \equiv \partial_\mu B'_\nu-\partial_\nu B'_\mu$) together with the respective generators $T^\alpha, T^a, {\mathcal Y}, B-L$ and gauge couplings $g_3, g_2, g_Y, g_1'$. Here $g_m$ takes into account the Abelian mixing between $U(1)_Y$ and $U(1)_{B-L}$. We do not propose this model as UV completion of the SM but just as an effective field theory valid up to the symmetry breaking scale $\chi_0$. 

As discussed around Eq.~(\ref{CWpot}), to realize RSB we  need the beta function of the quartic coupling of the flat-direction field $\chi$, in this case $\lambda_a$.
Using the general formalism of~\cite{Machacek:1983tz,Machacek:1983fi,Machacek:1984zw} we find the following one-loop expression
\be (4\pi)^2\mu\frac{d}{d\mu}\lambda_a = 96 g_1'^4 -48 \lambda_a g_1'^2+ 20\lambda_a^2+2\lambda_{ah}^2+2\lambda_a\Tr(yy^\dagger) - \Tr(yy^\dagger y y^\dagger).\ee
Evaluating now this beta function at a scale where $\lambda_a = 0$ to compute $\bar\beta$ defined in~(\ref{betabardef}) and neglecting $\lambda_{ah}$ and $y$ for the reasons explained above (all right-handed neutrino Majorana masses are taken below the EW scale in~\cite{Asaka:2005pn,Canetti:2012kh}) one finds 
\be \bar\beta = \frac{96 g_1'^4}{(4\pi)^2}. \ee 
One the other hand, in this setup the background-dependent mass of the new gauge boson, $Z'$, is 
\be m_{Z'}(\chi) = 2 |g_1'| \chi,\ee
where we used Eq.~(\ref{CovBmL}) and the fact that $|B-L|=2$ for the new scalar field $A$,
 the collective coupling $g$ defined in Eq.~(\ref{M2g2def}) is 
\be g = 2\sqrt{3} |g_1'|, \ee 
and $\tilde g$ defined in Eq.~(\ref{gtdef}) is 
\be \tilde g = 2\sqrt[3]{3} |g_1'|=\frac{g}{\sqrt[6]{3}}. \ee  Therefore, 
\be \bar\beta = \frac{2 g^4}{3(4\pi)^2}. \label{barbBmL} \ee
Like in the previous section, one can now use the model-independent analysis based on the improved supercool expansion of Sec.~\ref{Improved supercool expansion} with only two free parameters: $g$ and $\chi_0$.

Let us quantify the error that one is making in analysing this model with the standard supercool expansion at NLO of Sec.~\ref{recap}. Fig.~\ref{enloBL} (lower plots) shows $\epsilon$ and 
\be e_{nlo} \equiv \max(e_1,e_2,e_3), \qquad (e_1,e_2,e_3)\equiv \left(\frac{\tilde c_3^2 \tilde g^6 \epsilon}{2\pi^2 c_3^2 g^6}, \frac{|\frac12\log(\epsilon/g^2)-1/4|}{X},\frac{\epsilon}{3\times 32}\right) \label{enlo}\ee
(computed for simplicity with the LO formula for $T_n$ in~(\ref{appTn})). The estimate of the error  in~(\ref{enlo}) has been obtained like\footnote{In this case, however, one has an extra factor of $3$ in the denominator of the last entry because the  coefficient $\sum_b n_b m_b^4$ equals $g^4\chi^4/3$ now.}  in~(\ref{enlos}). As one can see, although $\epsilon$ is above 1 the quantity $e_{nlo}$ is small, especially for smaller values of $g$. The reason why this happens is again because we have more than one massive degrees of freedom. In this case, however, we have a single massive vector field, $Z'$, rather than two like in Sec.~\ref{A simple toy model} and so the suppression of the neglected terms is slightly weaker as one can see in Fig.~\ref{enloBL}. Fig.~\ref{enloBL} also shows that the improved supercool expansion is a better approximation than the standard supercool expansion at NLO as $e_1>e_2$ and $e_1> e_3$ in this case too. Again we attribute this to the fact that the number of degrees of freedom with dominant couplings with the flat-direction field $\chi$ is not very large (it is three here) and $\epsilon$ is not smaller than one in this case.

Since this model is phenomenologically very well motivated, it is also interesting to study the reheating after the supercooling period. In the following discussion we focus on the decay of the flat-direction field $\chi$ coming from $A$ into two physical Higgs bosons of mass $M_h\approx 125$~GeV. This process is induced by the portal interaction $\lambda_{ah} |A|^2|\mathcal{H}|^2$. Expanding around $\chi=\chi_0$ one obtains the effective interaction $\lambda_{ah}\chi_0\delta\chi |\mathcal{H}|^2$, where $\delta\chi \equiv\chi-\chi_0$.
On the other hand, using~(\ref{mchi}) and~(\ref{barbBmL}) one finds 
\be m_\chi = \sqrt{\frac{2}{3}}\frac{g^2}{4\pi} \chi_0. \ee
The radiative symmetry breaking of $B-L$ also induces EW symmetry breaking according the discussion around~(\ref{muhInd}) and a physical Higgs mass $M_h = \sqrt{\lambda_{ah}}\chi_0$. So the decay rate of $\chi$ in a pair of Higgs particles $\chi\to \mathcal{H}\mathcal{H}$ is (when $M_h$ is negligible compared to $m_\chi$)
\be \Gamma(\chi\to \mathcal{H}\mathcal{H}) = \frac{\lambda_{ah}^2\chi_0^2}{8\pi m_\chi} = \frac12 \sqrt{\frac32}\frac{M_h^4}{g^2\chi_0^3}.\ee 
The reheating temperature due to this channel may be computed through
\be T_r^4 = \frac{45 \Gamma^2(\chi\to \mathcal{H}\mathcal{H})\bp^2}{4\pi^3 g_*(T_r)}. \label{TRHBmL}\ee
But  this formula  is only valid if the radiation energy density $\rho_R$ does not exceed the vacuum energy density $\rho_V$ due to $\chi$ (because $\rho_V$ represents the full energy budget of the system). If this condition is not satisfied we determine $T_r$ as the maximal temperature compatible with $\rho_R\leq \rho_V$,  leading to the formula for $T_r$ in~(\ref{TRHmax}).
For $g$ of order one, $g_*\sim 10^2$ and
$\chi_0\lesssim 10^5$~GeV the reheating temperature is well above the EW scale and~(\ref{TRHmax}) holds such that the reheating effectively is fast. Increasing $\chi_0$ lowers $T_r$ in~(\ref{TRHBmL}) and $\rho_R\leq \rho_V$ can be satisfied.

 \section{Summary and conclusions}\label{Conclusions}
Let us conclude by providing a summary of the main original results obtained.
 \begin{itemize}
 \item In Sec.~\ref{impro} we have significantly extended the applicability of the model-independent approach to study PTs proposed in~\cite{Salvio:2023qgb}, which now works for a larger class of RSB models: the amount of supercooling required for the model-independent approach to work has been extended from $\epsilon$ small up to values of $\epsilon$ of order 1, where $\epsilon$ is defined in Eq.~(\ref{CondConv}).  \\
 First, in Sec.~\ref{Several degrees of freedom} it was pointed out that the supercool expansion proposed in~\cite{Salvio:2023qgb}  already gives an accurate model-independent description even for $\epsilon\sim1$ if there are several degrees of freedom with dominant couplings to the flat-direction field $\chi$ (the one responsible for RSB). \\
In Sec.~\ref{Improved supercool expansion} it was then explained how to improve the supercool expansion to obtain a good model-independent description for $\epsilon\sim1$ and an arbitrary number of (even few)  degrees of freedom with dominant couplings to $\chi$.   This has been achieved by including, unlike in~\cite{Salvio:2023qgb}, the cubic term of the effective potential in~(\ref{barVnlo}) in the non-perturbative computation of the bounce action and treating the other corrections as perturbations (indeed, those are still small as long as $\epsilon$ is at most of order one, as we discussed in Sec.~\ref{Several degrees of freedom}). Such ``improved supercool expansion" has been used to compute to good accuracy the nucleation temperature $T_n$ (and thus the strength $\alpha$ of the PT) as well as the inverse duration $\beta$ of the PT in terms of few parameters that are fixed once the model is specified: 
\begin{itemize}
\item $\chi_0$: the symmetry breaking scale 
\item $\bar\beta$: the beta function of the quartic coupling $\lambda_\chi$ of  $\chi$, evaluated at the scale where $\lambda_\chi$ vanishes. 
\item $g$: a sort of collective coupling of $\chi$ to all fields of the theory, which is precisely defined in Eq.~(\ref{M2g2def}). It is the square root of the sum of the squares of the couplings of $\chi$ to all fields.
\item $\tilde g$: an extra parameter that characterizes the size of the cubic term in~(\ref{barVnlo}). It is the cube root of the sum of the cubes of the couplings of $\chi$ to all bosonic fields, so $\tilde g\leq g$.
\end{itemize}  
 Analytical calculation can be performed to a greater extent in the approach of Sec.~\ref{Several degrees of freedom}, but the improved supercool expansion of Sec.~\ref{Improved supercool expansion} works for a larger class of models.
\item In Sec.~\ref{Applications} such improved supercool expansion has then been applied  to study in a model-independent way the spectrum of GWs  and the production of PBHs due to the first-order PT associated with the RSB. \\
 We have explained how to determine the GW spectrum (its amplitude and $f_{\rm peak}$) in terms of the above mentioned parameters in the hypothesis of fast reheating after supercooling. Among other things, we have found  values of $f_{\rm peak}$ and regions of the parameter space that correspond to the GW background recently detected by pulsar  timing arrays. Moreover, we have also found regions of the parameter space where the GW spectrum is above the sensitivity of LIGO-VIRGO O3 (which are then ruled out) and others that are within the reach of LISA. \\
 Furthermore, we have studied the generic validity of PBH production mechanisms in the RSB scenario for large supercooling. Also, we identified regions of the parameter space where PBHs produced by large over-densities due to an RSB PT can account for a significant  fraction of the dark matter density. Other mechanisms for PBH production can be active, however, in specific models. 
\item In Sec.~\ref{Examples of specific models}  we have applied the developed model-independent approach to study the PT in two RSB models: a simple illustrative one and a gauged $B-L$ phenomenological completion of the SM featuring right-handed neutrinos below the EW scale. In both these models there are more than one degrees of freedom with dominant couplings to $\chi$, but the number of these degrees of freedom is not much larger than one (it is six in the first model and three in the second one). Then we find that the improved supercool expansion of Sec.~\ref{Improved supercool expansion} works better than the method of Sec.~\ref{Several degrees of freedom}, which however already allows us to obtain a reasonably-good semi-analytical estimate of the PT properties. Given the phenomenological interest of the $B-L$ model, in the same section we have also studied reheating after supercooling, finding values of the parameters for which the reheating is fast and others for which it is not.
 \end{itemize}
 
 {\color{brown}

 }
 \vspace{0.2cm}

 \subsubsection*{Acknowledgments}
This work has been partially supported by the grant DyConn from the University of Rome Tor Vergata.

 \vspace{1cm}
\footnotesize
\begin{multicols}{2}

\end{multicols}


\begin{thebibliography}{}
\small{
 
 
  \bibitem{Abbott:2016blz} 
  B.~P.~Abbott {\it \textit{et al.}} [LIGO Scientific and Virgo Collaborations],
  ``Observation of Gravitational Waves from a Binary Black Hole Merger,''
  Phys.\ Rev.\ Lett.\  {\bf 116}, 061102 (2016)
  doi:10.1103/PhysRevLett.116.061102
  [\hhref{1602.03837}].

  
  
  
\bibitem{TheLIGOScientific:2016wyq1}
B.~Abbott \textit{\textit{et al.}} [LIGO Scientific and Virgo],
``GW150914: Implications for the stochastic gravitational wave background from binary black holes,''
Phys. Rev. Lett. \textbf{116}, no.13, 131102 (2016)
doi:10.1103/PhysRevLett.116.131102
[\hhref{1602.03847}].

\bibitem{LIGOScientific:2017ync}
B.~P.~Abbott \textit{et al.}
``Multi-messenger Observations of a Binary Neutron Star Merger,''
Astrophys. J. Lett. \textbf{848} (2017) no.2, L12
doi:10.3847/2041-8213/aa91c9
[\hhref{1710.05833}].
 
 
\bibitem{NANOGrav:2023gor}
G.~Agazie \textit{et al.} [NANOGrav],
``The NANOGrav 15 yr Data Set: Evidence for a Gravitational-wave Background,''
Astrophys. J. Lett. \textbf{951} (2023) no.1, L8
doi:10.3847/2041-8213/acdac6
[\hhref{2306.16213}].

\bibitem{Antoniadis:2023ott}
J.~Antoniadis, P.~Arumugam, S.~Arumugam, S.~Babak, M.~Bagchi, A.~S.~B.~Nielsen, C.~G.~Bassa, A.~Bathula, A.~Berthereau and M.~Bonetti, \textit{et al.}
``The second data release from the European Pulsar Timing Array III. Search for gravitational wave signals,''
[\hhref{2306.16214}].

\bibitem{Reardon:2023gzh}
D.~J.~Reardon, A.~Zic, R.~M.~Shannon, G.~B.~Hobbs, M.~Bailes, V.~Di Marco, A.~Kapur, A.~F.~Rogers, E.~Thrane and J.~Askew, \textit{et al.}
``Search for an Isotropic Gravitational-wave Background with the Parkes Pulsar Timing Array,''
Astrophys. J. Lett. \textbf{951} (2023) no.1, L6
doi:10.3847/2041-8213/acdd02
[\hhref{2306.16215}].

\bibitem{Xu:2023wog}
H.~Xu, S.~Chen, Y.~Guo, J.~Jiang, B.~Wang, J.~Xu, Z.~Xue, R.~N.~Caballero, J.~Yuan and Y.~Xu, \textit{et al.}
``Searching for the Nano-Hertz Stochastic Gravitational Wave Background with the Chinese Pulsar Timing Array Data Release I,''
Res. Astron. Astrophys. \textbf{23} (2023) no.7, 075024
doi:10.1088/1674-4527/acdfa5
[\hhref{2306.16216}].


\bibitem{Hawking:1982ga}
S.~W.~Hawking, I.~G.~Moss and J.~M.~Stewart,
``Bubble Collisions in the Very Early Universe,''
Phys. Rev. D \textbf{26} (1982), 2681
doi:10.1103/PhysRevD.26.2681

\bibitem{Moss:1994iq}
I.~G.~Moss,
``Singularity formation from colliding bubbles,''
Phys. Rev. D \textbf{50} (1994), 676-681
doi:10.1103/PhysRevD.50.676

\bibitem{Crawford:1982yz}
M.~Crawford and D.~N.~Schramm,
``Spontaneous Generation of Density Perturbations in the Early Universe,''
Nature \textbf{298} (1982), 538-540
doi:10.1038/298538a0

\bibitem{Gross:2021qgx}
C.~Gross, G.~Landini, A.~Strumia and D.~Teresi,
``Dark Matter as dark dwarfs and other macroscopic objects: multiverse relics?,''
JHEP \textbf{09} (2021), 033
doi:10.1007/JHEP09(2021)033
[\hhref{2105.02840}].

\bibitem{Baker:2021sno}
M.~J.~Baker, M.~Breitbach, J.~Kopp and L.~Mittnacht,
``Detailed Calculation of Primordial Black Hole Formation During First-Order Cosmological Phase Transitions,''
[\hhref{2110.00005}].

\bibitem{Jung:2021mku}
T.~H.~Jung and T.~Okui,
``Primordial black holes from bubble collisions during a first-order phase transition,''
[\hhref{2110.04271}].

\bibitem{Kawana:2021tde}
K.~Kawana and K.~P.~Xie,
``Primordial black holes from a cosmic phase transition: The collapse of Fermi-balls,''
Phys. Lett. B \textbf{824} (2022), 136791
doi:10.1016/j.physletb.2021.136791
[\hhref{2106.00111}].

\bibitem{Kodama:1982sf}
H.~Kodama, M.~Sasaki and K.~Sato,
``Abundance of Primordial Holes Produced by Cosmological First Order Phase Transition,''
Prog. Theor. Phys. \textbf{68} (1982), 1979
doi:10.1143/PTP.68.1979

\bibitem{Liu:2021svg}
J.~Liu, L.~Bian, R.~G.~Cai, Z.~K.~Guo and S.~J.~Wang,
``Primordial black hole production during first-order phase transitions,''
Phys. Rev. D \textbf{105} (2022) no.2, L021303
doi:10.1103/PhysRevD.105.L021303
[\hhref{2106.05637}].
%
\bibitem{Hashino:2021qoq}
K.~Hashino, S.~Kanemura and T.~Takahashi,
``Primordial black holes as a probe of strongly first-order electroweak phase transition,''
Phys. Lett. B \textbf{833} (2022), 137261
doi:10.1016/j.physletb.2022.137261
[\hhref{2111.13099}].

\bibitem{Kawana:2022olo}
K.~Kawana, T.~Kim and P.~Lu,
``PBH Formation from Overdensities in Delayed Vacuum Transitions,''
[\hhref{2212.14037}].

\bibitem{Gouttenoire:2023naa}
Y.~Gouttenoire and T.~Volansky,
``Primordial Black Holes from Supercooled Phase Transitions,''
[\hhref{2305.04942}].

\bibitem{Salvio:2023qgb}
A.~Salvio,
``Model-independent radiative symmetry breaking and gravitational waves,''
JCAP \textbf{04} (2023), 051
doi:10.1088/1475-7516/2023/04/051
[\hhref{2302.10212}].

\bibitem{Coleman:1973jx}
S.~R.~Coleman and E.~J.~Weinberg,
``Radiative Corrections as the Origin of Spontaneous Symmetry Breaking,''
Phys. Rev. D \textbf{7} (1973), 1888-1910
doi:10.1103/PhysRevD.7.1888

\bibitem{Levi:2022bzt}
N.~Levi, T.~Opferkuch and D.~Redigolo,
``The supercooling window at weak and strong coupling,''
JHEP \textbf{02}, 125 (2023)
doi:10.1007/JHEP02(2023)125
[\hhref{2212.08085}].



\bibitem{Gildener:1976ih}
E.~Gildener and S.~Weinberg,
``Symmetry Breaking and Scalar Bosons,''
Phys. Rev. D \textbf{13} (1976), 3333
doi:10.1103/PhysRevD.13.3333


\bibitem{Witten:1980ez}
E.~Witten,
``Cosmological Consequences of a Light Higgs Boson,''
Nucl. Phys. B \textbf{177} (1981), 477-488
doi:10.1016/0550-3213(81)90182-6



\bibitem{Espinosa:2008kw}
J.~R.~Espinosa, T.~Konstandin, J.~M.~No and M.~Quiros,
``Some Cosmological Implications of Hidden Sectors,''
Phys. Rev. D \textbf{78} (2008), 123528
doi:10.1103/PhysRevD.78.123528
[\hhref{0809.3215}].

\bibitem{Farzinnia:2014yqa}
A.~Farzinnia and J.~Ren,
``Strongly First-Order Electroweak Phase Transition and Classical Scale Invariance,''
Phys. Rev. D \textbf{90} (2014) no.7, 075012
doi:10.1103/PhysRevD.90.075012
[\hhref{1408.3533}].

\bibitem{Sannino:2015wka}
F.~Sannino and J.~Virkaj\"arvi,
``First Order Electroweak Phase Transition from (Non)Conformal Extensions of the Standard Model,''
Phys. Rev. D \textbf{92} (2015) no.4, 045015 doi:10.1103/PhysRevD.92.045015
[\hhref{1505.05872}].

\bibitem{Marzola:2017jzl}
L.~Marzola, A.~Racioppi and V.~Vaskonen,
``Phase transition and gravitational wave phenomenology of scalar conformal extensions of the Standard Model,''
Eur. Phys. J. C \textbf{77} (2017) no.7, 484 doi:10.1140/epjc/s10052-017-4996-1
[\hhref{1704.01034}].

\bibitem{Brdar:2019qut}
V.~Brdar, A.~J.~Helmboldt and M.~Lindner,
``Strong Supercooling as a Consequence of Renormalization Group Consistency,''
JHEP \textbf{12} (2019), 158
doi:10.1007/JHEP12(2019)158
[\hhref{1910.13460}].

\bibitem{Kierkla:2022odc}
M.~Kierkla, A.~Karam and B.~Swiezewska,
``Conformal model for gravitational waves and dark matter: A status update,'' JHEP \textbf{03} (2023), 007 doi:10.1007/JHEP03(2023)007
[\hhref{2210.07075}].

\bibitem{Huang:2020bbe}
W.~C.~Huang, F.~Sannino and Z.~W.~Wang,
``Gravitational Waves from Pati-Salam Dynamics,''
Phys. Rev. D \textbf{102}, no.9, 095025 (2020) doi:10.1103/PhysRevD.102.095025
[\hhref{2004.02332}].

\bibitem{Peccei:1977hh}
R.~D.~Peccei and H.~R.~Quinn,
``CP Conservation in the Presence of Instantons,''
Phys. Rev. Lett. \textbf{38} (1977), 1440-1443
doi:10.1103/PhysRevLett.38.1440.
R.~D.~Peccei and H.~R.~Quinn,
  ``Constraints Imposed by CP Conservation in the Presence of Instantons,''
  Phys.\ Rev.\ D {\bf 16} (1977) 1791  doi:10.1103/PhysRevD.16.1791

\bibitem{DelleRose:2019pgi}
L.~Delle Rose, G.~Panico, M.~Redi and A.~Tesi,
``Gravitational Waves from Supercool Axions,''
JHEP \textbf{04} (2020), 025 doi:10.1007/JHEP04(2020)025
[\hhref{1912.06139}].

\bibitem{VonHarling:2019rgb}
B.~Von Harling, A.~Pomarol, O.~Pujol\`as and F.~Rompineve,
``Peccei-Quinn Phase Transition at LIGO,''
JHEP \textbf{04} (2020), 195
doi:10.1007/JHEP04(2020)195
[\hhref{1912.07587}].

\bibitem{Salvio:2020prd}
A.~Salvio,
``A fundamental QCD axion model,''
Phys. Lett. B \textbf{808} (2020), 135686
doi:10.1016/j.physletb.2020.135686
[\hhref{2003.10446}].

\bibitem{Ghoshal:2020vud}
A.~Ghoshal and A.~Salvio,
``Gravitational waves from fundamental axion dynamics,''
JHEP \textbf{12} (2020), 049
doi:10.1007/JHEP12(2020)049
[\hhref{2007.00005}].

\bibitem{Brdar:2018num}
V.~Brdar, A.~J.~Helmboldt and J.~Kubo,
``Gravitational Waves from First-Order Phase Transitions: LIGO as a Window to Unexplored Seesaw Scales,''
JCAP \textbf{02} (2019), 021
doi:10.1088/1475-7516/2019/02/021
[\hhref{1810.12306}].

\bibitem{Kubo:2020fdd}
J.~Kubo, J.~Kuntz, M.~Lindner, J.~Rezacek, P.~Saake and A.~Trautner,
``Unified emergence of energy scales and cosmic inflation,''
JHEP \textbf{08} (2021), 016
doi:10.1007/JHEP08(2021)016
[\hhref{2012.09706}].

\bibitem{Salvio:2020axm}
A.~Salvio,
``Dimensional Transmutation in Gravity and Cosmology,''
Int. J. Mod. Phys. A \textbf{36} (2021) no.08n09, 2130006
doi:10.1142/S0217751X21300064
[\hhref{2012.11608}].


\bibitem{KAGRA:2021kbb}
R.~Abbott \textit{et al.} [KAGRA, Virgo and LIGO Scientific],
``Upper limits on the isotropic gravitational-wave background from Advanced LIGO and Advanced Virgo\textquoteright{}s third observing run,''
Phys. Rev. D \textbf{104} (2021) no.2, 022004
doi:10.1103/PhysRevD.104.022004
[\hhref{2101.12130}].




\bibitem{Harry:2010zz}
G.~M.~Harry [LIGO Scientific],
``Advanced LIGO: The next generation of GW detectors,''
Class. Quant. Grav. \textbf{27}, 084006 (2010)
doi:10.1088/0264-9381/27/8/084006

\bibitem{TheLIGOScientific:2014jea}
J.~Aasi \textit{\textit{et al.}} [LIGO Scientific],
``Advanced LIGO,''
Class. Quant. Grav. \textbf{32}, 074001 (2015)
doi:10.1088/0264-9381/32/7/074001
[\hhref{1411.4547}].

\bibitem{VIRGO:2014yos}
F.~Acernese \textit{et al.} [VIRGO],
``Advanced Virgo: a second-generation interferometric gravitational wave detector,''
Class. Quant. Grav. \textbf{32} (2015) no.2, 024001
doi:10.1088/0264-9381/32/2/024001
[\hhref{1408.3978}].



\bibitem{Audley:2017drz}
P.~Amaro-Seoane \textit{\textit{et al.}} [LISA],
``Laser Interferometer Space Antenna,''
[\hhref{1702.00786}].

\bibitem{Evans:2016mbw}
B.~P.~Abbott \textit{\textit{et al.}} [LIGO Scientific],
``Exploring the Sensitivity of Next Generation Gravitational Wave Detectors,''
Class. Quant. Grav. \textbf{34}, no.4, 044001 (2017)
doi:10.1088/1361-6382/aa51f4
[\hhref{1607.08697}].


\bibitem{Reitze:2019iox}
D.~Reitze \textit{et al.},
``Cosmic Explorer: The U.S. Contribution to Gravitational-Wave Astronomy beyond LIGO,''
Bull. Am. Astron. Soc. \textbf{51}, 035
[\hhref{1907.04833}].

\bibitem{Punturo:2010zz}
M.~Punturo \textit{et al.},
``The Einstein Telescope: A third-generation GW observatory,''
Class. Quant. Grav. \textbf{27}, 194002 (2010)
doi:10.1088/0264-9381/27/19/194002

\bibitem{Hild:2010id}
S.~Hild \textit{et al.},
``Sensitivity Studies for Third-Generation Gravitational Wave Observatories,''
Class. Quant. Grav. \textbf{28}, 094013 (2011) doi:10.1088/0264-9381/28/9/094013
[\hhref{1012.0908}].

\bibitem{Sathyaprakash:2012jk}
B.~Sathyaprakash\textit{et al.},
``Scientific Objectives of Einstein Telescope,''
Class. Quant. Grav. \textbf{29}, 124013 (2012)
doi:10.1088/0264-9381/29/12/124013
[\hhref{1206.0331}].


\bibitem{Crowder:2005nr}
J.~Crowder and N.~J.~Cornish,
``Beyond LISA: Exploring future GW missions,''
Phys. Rev. D \textbf{72}, 083005 (2005)
doi:10.1103/PhysRevD.72.083005
[\hhref{gr-qc/0506015}].


\bibitem{Corbin:2005ny}
V.~Corbin and N.~J.~Cornish,
``Detecting the cosmic GW background with the big bang observer,''
Class. Quant. Grav. \textbf{23}, 2435-2446 (2006)
doi:10.1088/0264-9381/23/7/014
[\hhref{gr-qc/0512039}].

\bibitem{Harry:2006fi}
G.~Harry, P.~Fritschel, D.~Shaddock, W.~Folkner and E.~Phinney,
``Laser interferometry for the big bang observer,''
Class. Quant. Grav. \textbf{23}, 4887-4894 (2006)
doi:10.1088/0264-9381/23/15/008
[\hhref{gr-qc/0506015}]

\bibitem{Seto:2001qf}
N.~Seto, S.~Kawamura and T.~Nakamura,
``Possibility of direct measurement of the acceleration of the universe using 0.1-Hz band laser interferometer GW antenna in space,''
Phys. Rev. Lett. \textbf{87}, 221103 (2001)
doi:10.1103/PhysRevLett.87.221103
[\hhref{astro-ph/0108011}].

\bibitem{Kawamura:2006}
        S.~Kawamura \textit{\textit{et al.}}, ``The Japanese space gravitational wave antenna - DECIGO," 
Class. Quant. Grav. \textbf{23} (2006), S125-S132.
doi:10.1088/0264-9381/23/8/S17

\bibitem{Salvio:2014soa}
A.~Salvio and A.~Strumia,
``Agravity,''
JHEP \textbf{06} (2014), 080
doi:10.1007/JHEP06(2014)080
[\hhref{1403.4226}].

\bibitem{Kannike:2015apa}
K.~Kannike, G.~H\"utsi, L.~Pizza, A.~Racioppi, M.~Raidal, A.~Salvio and A.~Strumia,
``Dynamically Induced Planck Scale and Inflation,''
JHEP \textbf{05} (2015), 065
doi:10.1007/JHEP05(2015)065
[\hhref{1502.01334}].

\bibitem{Salvio:2017qkx}
A.~Salvio and A.~Strumia,
``Agravity up to infinite energy,''
Eur. Phys. J. C \textbf{78} (2018) no.2, 124
doi:10.1140/epjc/s10052-018-5588-4
[\hhref{1705.03896}].

\bibitem{Salvio:2017xul}
A.~Salvio,
``Inflationary Perturbations in No-Scale Theories,''
Eur. Phys. J. C \textbf{77} (2017) no.4, 267
doi:10.1140/epjc/s10052-017-4825-6
[\hhref{1703.08012}].

\bibitem{Salvio:2019wcp}
A.~Salvio,
``Quasi-Conformal Models and the Early Universe,''
Eur. Phys. J. C \textbf{79} (2019) no.9, 750
doi:10.1140/epjc/s10052-019-7267-5
[\hhref{1907.00983}].

\bibitem{Alvarez-Luna:2022hka}
C.~\'Alvarez-Luna, S.~de la Calle-Leal, J.~A.~R.~Cembranos and J.~J.~Sanz-Cillero,
``Gravitational Coleman-Weinberg mechanism,''
JHEP \textbf{02} (2023), 232
doi:10.1007/JHEP02(2023)232
[\hhref{2212.01785}].

\bibitem{Arnold:1992rz}
P.~B.~Arnold and O.~Espinosa,
``The Effective potential and first order phase transitions: Beyond leading-order,''
Phys. Rev. D \textbf{47} (1993), 3546
[erratum: Phys. Rev. D \textbf{50} (1994), 6662]
doi:10.1103/PhysRevD.47.3546
[\hhref{hep-ph/9212235}].

\bibitem{Kierkla:2023von}
M.~Kierkla, B.~Swiezewska, T.~V.~I.~Tenkanen and J.~van de Vis,
``Gravitational waves from supercooled phase transitions: dimensional transmutation meets dimensional reduction,''
[\hhref{2312.12413}].

\bibitem{Dolan:1973qd}
L.~Dolan and R.~Jackiw,
``Symmetry Behavior at Finite Temperature,''
Phys. Rev. D \textbf{9} (1974), 3320-3341
doi:10.1103/PhysRevD.9.3320


\bibitem{Coleman:1977py}
S.~R.~Coleman,
``The Fate of the False Vacuum. 1. Semiclassical Theory,''
Phys. Rev. D \textbf{15} (1977), 2929-2936
[erratum: Phys. Rev. D \textbf{16} (1977), 1248]
doi:10.1103/PhysRevD.16.1248

\bibitem{Callan:1977pt}
C.~G.~Callan, Jr. and S.~R.~Coleman,
``The Fate of the False Vacuum. 2. First Quantum Corrections,''
Phys. Rev. D \textbf{16} (1977), 1762-1768 doi:10.1103/PhysRevD.16.1762

\bibitem{Linde:1980tt}
A.~D.~Linde,
``Fate of the False Vacuum at Finite Temperature: Theory and Applications,''
Phys. Lett. B \textbf{100} (1981), 37-40 doi:10.1016/0370-2693(81)90281-1

\bibitem{Linde:1981zj}
A.~D.~Linde,
``Decay of the False Vacuum at Finite Temperature,''
Nucl. Phys. B \textbf{216} (1983), 421. Erratum: [Nucl. Phys. B \textbf{223}, 544 (1983)] doi:10.1016/0550-3213(83)90072-X


\bibitem{Salvio:2016mvj}
A.~Salvio, A.~Strumia, N.~Tetradis and A.~Urbano,
``On gravitational and thermal corrections to vacuum decay,''
JHEP \textbf{09} (2016), 054
doi:10.1007/JHEP09(2016)054
[\hhref{1608.02555}].


\bibitem{Brezin-Parisi} E. Brezin and G. Parisi, J. Stat. Phys. 19 (1978) 269 doi:10.1007/BF01011726

\bibitem{Arnold:1991cv}
P.~B.~Arnold and S.~Vokos,
``Instability of hot electroweak theory: bounds on m(H) and M(t),''
Phys. Rev. D \textbf{44} (1991), 3620-3627
doi:10.1103/PhysRevD.44.3620


\bibitem{Caprini:2019egz}
C.~Caprini, M.~Chala, G.~C.~Dorsch, M.~Hindmarsh, S.~J.~Huber, T.~Konstandin, J.~Kozaczuk, G.~Nardini, J.~M.~No and K.~Rummukainen, \textit{et al.}
``Detecting gravitational waves from cosmological phase transitions with LISA: an update,''
JCAP \textbf{03} (2020), 024
doi:10.1088/1475-7516/2020/03/024
[\hhref{1910.13125}].

\bibitem{Ellis:2019oqb}
J.~Ellis, M.~Lewicki, J.~M.~No and V.~Vaskonen,
``Gravitational wave energy budget in strongly supercooled phase transitions,''
JCAP \textbf{06} (2019), 024
doi:10.1088/1475-7516/2019/06/024
[\hhref{1903.09642}].

\bibitem{Caprini:2015zlo}
C.~Caprini, M.~Hindmarsh, S.~Huber, T.~Konstandin, J.~Kozaczuk, G.~Nardini, J.~M.~No, A.~Petiteau, P.~Schwaller, G.~Servant and D.~J.~Weir,
``Science with the space-based interferometer eLISA. II: GWs from cosmological phase transitions,''
JCAP \textbf{04} (2016), 001
doi:10.1088/1475-7516/2016/04/001
[\hhref{1512.06239}].

\bibitem{Caprini:2018mtu}
C.~Caprini and D.~G.~Figueroa,
``Cosmological Backgrounds of Gravitational Waves,''
Class. Quant. Grav. \textbf{35} (2018) no.16, 163001
doi:10.1088/1361-6382/aac608
[\hhref{1801.04268}].




\bibitem{Quiros:1994dr}
  M.~Quiros,
  ``Field theory at finite temperature and phase transitions,''
  Helv.\ Phys.\ Acta {\bf 67} (1994) 451 doi:10.5169/seals-116659. 
  
  
\bibitem{Adams:1993zs}
F.~C.~Adams,
``General solutions for tunneling of scalar fields with quartic potentials,''
Phys. Rev. D \textbf{48}, 2800-2805 (1993)
doi:10.1103/PhysRevD.48.2800
[\hhref{hep-ph/9302321}].

\bibitem{Sarid:1998sn}
U.~Sarid,
``Tools for tunneling,''
Phys. Rev. D \textbf{58}, 085017 (1998)
doi:10.1103/PhysRevD.58.085017
[\hhref{hep-ph/9804308}].

\bibitem{dataset}
A.~Salvio, ``Supercooling in Radiative Symmetry Breaking: Theory Extensions, Gravitational Wave Detection and Primordial Black Holes; Dataset,"   (2023) https://doi.org/10.5281/zenodo.8128176

\bibitem{Bodeker:2009qy}
D.~Bodeker and G.~D.~Moore,
``Can electroweak bubble walls run away?,''
JCAP \textbf{05} (2009), 009
doi:10.1088/1475-7516/2009/05/009
[\hhref{0903.4099}].

\bibitem{Bodeker:2017cim}
D.~Bodeker and G.~D.~Moore,
``Electroweak Bubble Wall Speed Limit,''
JCAP \textbf{05} (2017), 025
doi:10.1088/1475-7516/2017/05/025
[\hhref{1703.08215}].

\bibitem{Kearney:2015vba}
J.~Kearney, H.~Yoo and K.~M.~Zurek,
``Is a Higgs Vacuum Instability Fatal for High-Scale Inflation?,''
Phys. Rev. D \textbf{91} (2015) no.12, 123537
doi:10.1103/PhysRevD.91.123537
[\hhref{1503.05193}].

\bibitem{Joti:2017fwe}
A.~Joti, A.~Katsis, D.~Loupas, A.~Salvio, A.~Strumia, N.~Tetradis and A.~Urbano,
``(Higgs) vacuum decay during inflation,''
JHEP \textbf{07} (2017), 058
doi:10.1007/JHEP07(2017)058
[\hhref{1706.00792}].

\bibitem{Markkanen:2018pdo}
T.~Markkanen, A.~Rajantie and S.~Stopyra,
``Cosmological Aspects of Higgs Vacuum Metastability,''
Front. Astron. Space Sci. \textbf{5} (2018), 40
doi:10.3389/fspas.2018.00040
[\hhref{1809.06923}].


\bibitem{Maggiore:2018sht}
M.~Maggiore,
``Gravitational Waves. Vol. 2: Astrophysics and Cosmology,'' Oxford University Press, 3, 2018.

\bibitem{Lewicki:2022pdb}
M.~Lewicki and V.~Vaskonen,
``Gravitational waves from bubble collisions and fluid motion in strongly supercooled phase transitions,''
Eur. Phys. J. C \textbf{83} (2023) no.2, 109
doi:10.1140/epjc/s10052-023-11241-3
[\hhref{2208.11697}].



\bibitem{Huber:2008hg}
S.~J.~Huber and T.~Konstandin,
``Gravitational Wave Production by Collisions: More Bubbles,''
JCAP \textbf{09} (2008), 022
doi:10.1088/1475-7516/2008/09/022
[\hhref{0806.1828}].



\bibitem{Kosowsky:1991ua}
A.~Kosowsky, M.~S.~Turner and R.~Watkins,
``Gravitational radiation from colliding vacuum bubbles,''
Phys. Rev. D \textbf{45}, 4514-4535 (1992)
doi:10.1103/PhysRevD.45.4514

\bibitem{Kosowsky:1992rz}
A.~Kosowsky, M.~S.~Turner and R.~Watkins,
``Gravitational waves from first order cosmological phase transitions,''
Phys. Rev. Lett. \textbf{69}, 2026-2029 (1992)
doi:10.1103/PhysRevLett.69.2026



\bibitem{Jinno:2017fby}
R.~Jinno and M.~Takimoto,
``Gravitational waves from bubble dynamics: Beyond the Envelope,''
JCAP \textbf{01} (2019), 060
doi:10.1088/1475-7516/2019/01/060
[\hhref{1707.03111}].

\bibitem{Konstandin:2017sat}
T.~Konstandin,
``Gravitational radiation from a bulk flow model,''
JCAP \textbf{03} (2018), 047
doi:10.1088/1475-7516/2018/03/047
[\hhref{1712.06869}].

\bibitem{Lewicki:2019gmv}
M.~Lewicki and V.~Vaskonen,
``On bubble collisions in strongly supercooled phase transitions,''
Phys. Dark Univ. \textbf{30} (2020), 100672
doi:10.1016/j.dark.2020.100672
[\hhref{1912.00997}].

\bibitem{Lewicki:2020jiv}
M.~Lewicki and V.~Vaskonen,
``Gravitational wave spectra from strongly supercooled phase transitions,''
Eur. Phys. J. C \textbf{80} (2020) no.11, 1003
doi:10.1140/epjc/s10052-020-08589-1
[\hhref{2007.04967}].

\bibitem{Lewicki:2020azd}
M.~Lewicki and V.~Vaskonen,
``Gravitational waves from colliding vacuum bubbles in gauge theories,''
Eur. Phys. J. C \textbf{81} (2021) no.5, 437
[erratum: Eur. Phys. J. C \textbf{81} (2021) no.12, 1077]
doi:10.1140/epjc/s10052-021-09232-3
[\hhref{2012.07826}].



\bibitem{Freese:2022qrl}
K.~Freese and M.~W.~Winkler,
``Have pulsar timing arrays detected the hot big bang: Gravitational waves from strong first order phase transitions in the early Universe,''
Phys. Rev. D \textbf{106} (2022) no.10, 103523
doi:10.1103/PhysRevD.106.103523
[\hhref{2208.03330}].



\bibitem{Fabbrichesi:2020wbt}
M.~Fabbrichesi, E.~Gabrielli and G.~Lanfranchi,
``The Dark Photon,''
doi:10.1007/978-3-030-62519-1
[\hhref{2005.01515}].


\bibitem{NANOGrav:2023hvm}
A.~Afzal \textit{et al.} [NANOGrav],
``The NANOGrav 15 yr Data Set: Search for Signals from New Physics,''
Astrophys. J. Lett. \textbf{951} (2023) no.1, L11
doi:10.3847/2041-8213/acdc91
[\hhref{2306.16219}].

\bibitem{Antoniadis:2023zhi}
J.~Antoniadis, P.~Arumugam, S.~Arumugam, P.~Auclair, S.~Babak, M.~Bagchi, A.~S.~Bak Nielsen, E.~Barausse, C.~G.~Bassa and A.~Bathula, \textit{et al.}
``The second data release from the European Pulsar Timing Array: V. Implications for massive black holes, dark matter and the early Universe,''
[\hhref{2306.16227}].

\bibitem{Bringmann:2023opz}
T.~Bringmann, P.~F.~Depta, T.~Konstandin, K.~Schmidt-Hoberg and C.~Tasillo,
``Does NANOGrav observe a dark sector phase transition?,''
[\hhref{2306.09411}].

\bibitem{Madge:2023cak}
E.~Madge, E.~Morgante, C.~P.~Ib\'a\~nez, N.~Ramberg and S.~Schenk,
``Primordial gravitational waves in the nano-Hertz regime and PTA data -- towards solving the GW inverse problem,''
[\hhref{2306.14856}].

\bibitem{Zu:2023olm}
L.~Zu, C.~Zhang, Y.~Y.~Li, Y.~C.~Gu, Y.~L.~S.~Tsai and Y.~Z.~Fan,
``Mirror QCD phase transition as the origin of the nanohertz Stochastic Gravitational-Wave Background detected by the Pulsar Timing Arrays,''
[\hhref{2306.16769}].

\bibitem{Han:2023olf}
C.~Han, K.~P.~Xie, J.~M.~Yang and M.~Zhang,
``Self-interacting dark matter implied by nano-Hertz gravitational waves,''
[\hhref{2306.16966}].

\bibitem{Fujikura:2023lkn}
K.~Fujikura, S.~Girmohanta, Y.~Nakai and M.~Suzuki,
``NANOGrav Signal from a Dark Conformal Phase Transition,''
[\hhref{2306.17086}].

\bibitem{Kitajima:2023cek}
N.~Kitajima, J.~Lee, K.~Murai, F.~Takahashi and W.~Yin,
``Nanohertz Gravitational Waves from Axion Domain Walls Coupled to QCD,''
[\hhref{2306.17146}].

\bibitem{Bai:2023cqj}
Y.~Bai, T.~K.~Chen and M.~Korwar,
``QCD-Collapsed Domain Walls: QCD Phase Transition and Gravitational Wave Spectroscopy,''
[\hhref{2306.17160}].

\bibitem{Addazi:2023jvg}
A.~Addazi, Y.~F.~Cai, A.~Marciano and L.~Visinelli,
``Have pulsar timing array methods detected a cosmological phase transition?,''
[\hhref{2306.17205}].

\bibitem{Athron:2023mer}
P.~Athron, A.~Fowlie, C.~T.~Lu, L.~Morris, L.~Wu, Y.~Wu and Z.~Xu,
``Can Supercooled Phase Transitions explain the Gravitational Wave Background observed by Pulsar Timing Arrays?,''
[\hhref{2306.17239}].

\bibitem{Lu:2023mcz}
B.~Q.~Lu and C.~W.~Chiang,
``Nano-Hertz stochastic gravitational wave background from domain wall annihilation,''
[\hhref{2307.00746}].

\bibitem{Xiao:2023dbb}
Y.~Xiao, J.~M.~Yang and Y.~Zhang,
``Implications of Nano-Hertz Gravitational Waves on Electroweak Phase Transition in the Singlet Dark Matter Model,''
[\hhref{2307.01072}].

\bibitem{Li:2023bxy}
S.~P.~Li and K.~P.~Xie,
``A collider test of nano-Hertz gravitational waves from pulsar timing arrays,''
[\hhref{2307.01086}].

\bibitem{Ghosh:2023aum}
T.~Ghosh, A.~Ghoshal, H.~K.~Guo, F.~Hajkarim, S.~F.~King, K.~Sinha, X.~Wang and G.~White,
``Did we hear the sound of the Universe boiling? Analysis using the full fluid velocity profiles and NANOGrav 15-year data,''
[\hhref{2307.02259}].

\bibitem{Wu:2023hsa}
Y.~M.~Wu, Z.~C.~Chen and Q.~G.~Huang,
``Cosmological Interpretation for the Stochastic Signal in Pulsar Timing Arrays,''
[\hhref{2307.03141}].

\bibitem{DiBari:2023upq}
P.~Di Bari and M.~H.~Rahat,
``The split majoron model confronts the NANOGrav signal,''
[\hhref{2307.03184}].


\bibitem{Babak:2021mhe}
S.~Babak, A.~Petiteau and M.~Hewitson,
``LISA Sensitivity and SNR Calculations,''
[\hhref{2108.01167}].


\bibitem{Pati:1974yy}
J.~C.~Pati and A.~Salam,
``Lepton Number as the Fourth Color,''
Phys. Rev. D \textbf{10} (1974), 275-289
[erratum: Phys. Rev. D \textbf{11} (1975), 703-703]
doi:10.1103/PhysRevD.10.275

\bibitem{Babu:1985gi}
K.~S.~Babu, X.~G.~He and S.~Pakvasa,
``Neutrino Masses and Proton Decay Modes in SU(3) X SU(3) X SU(3) Trinification,''
Phys. Rev. D \textbf{33} (1986), 763
doi:10.1103/PhysRevD.33.763


\bibitem{Asaka:2005pn}
  T.~Asaka and M.~Shaposhnikov,
  ``The nuMSM, dark matter and baryon asymmetry of the universe,''
  Phys.\ Lett.\ B {\bf 620} (2005) 17
doi:10.1016/j.physletb.2005.06.020
  [\hhref{hep-ph/0505013}].
  
\bibitem{Canetti:2012kh}
  L.~Canetti, M.~Drewes, T.~Frossard and M.~Shaposhnikov,
  ``Dark Matter, Baryogenesis and Neutrino Oscillations from Right Handed Neutrinos,''
  Phys.\ Rev.\ D {\bf 87} (2013) 093006
  doi:10.1103/PhysRevD.87.093006
   [\hhref{1208.4607}].


\bibitem{Iso:2009ss}
S.~Iso, N.~Okada and Y.~Orikasa,
``Classically conformal $B^-$ L extended Standard Model,''
Phys. Lett. B \textbf{676}, 81-87 (2009)
doi:10.1016/j.physletb.2009.04.046
[\hhref{0902.4050}].


\bibitem{Machacek:1983tz}
  M.~E.~Machacek and M.~T.~Vaughn,
  ``Two Loop Renormalization Group Equations in a General Quantum Field Theory. 1. Wave Function Renormalization,''
  Nucl.\ Phys.\ B {\bf 222} (1983) 83
  doi:10.1016/0550-3213(83)90610-7
  
\bibitem{Machacek:1983fi}
  M.~E.~Machacek and M.~T.~Vaughn,
  ``Two Loop Renormalization Group Equations in a General Quantum Field Theory. 2. Yukawa Couplings,''
  Nucl.\ Phys.\ B {\bf 236} (1984) 221
  doi:10.1016/0550-3213(84)90533-9
  
\bibitem{Machacek:1984zw}
  M.~E.~Machacek and M.~T.~Vaughn,
  ``Two Loop Renormalization Group Equations in a General Quantum Field Theory. 3. Scalar Quartic Couplings,''
  Nucl.\ Phys.\ B {\bf 249} (1985) 70
  doi:10.1016/0550-3213(85)90040-9


 



  
  }
  

\end{thebibliography}
\end{document}